# Notes on "Notes on the Synthesis of Form"

## Dawning Insights in Early Christopher Alexander

Richard P. Gabriel°

- Hasso Plattner Institute, University of Potsdam, Germany (Visiting Researcher)

**Abstract**
This essay is a picaresque—a first-person narrative relating the adventures of a rogue (me) sifting through the mind of Christopher Alexander as he left behind formalized design thinking in favor of a more intuitive, almost spiritual process.

The work of Christopher Alexander is familiar to many computer scientists: for some it's patterns, for some it's the mystical *quality without a name* and "Nature of Order"; for many more it's "Notes on the Synthesis of Form"—Alexander's formalized design method and foreshadowing ideas about cohesion and coupling in software. Since the publication of "Design Patterns" by Gamma et al. in 1994, there have been hundreds of books published about design / software patterns, thousands of published pattern languages, and tens of thousands of published patterns.

"Notes," published in 1964, was quickly followed by one of Alexander's most important essays, "A City is Not a Tree," in which he repudiates the formal method described in "Notes," and his Preface to the paperback edition of "Notes" in 1971 repeats the repudiation. For many close readers of Alexander, this discontinuity is startling and unexplained.

When I finally read "Notes" in 2015, I was struck by the detailed worked example, along with a peculiar mathematical treatment of the method, and a hint that the modularization presented in the example was reckoned by a computer program he had written—all in the late 1950s and early 1960s. Because of my fascination with metaheuristic optimization, I couldn't resist trying to replicate his experimental results.

Computers and their programs relish dwelling on flaws in your thinking—Alexander was not exempt. By engaging in hermeneutics and software archeology, I was able to uncover / discover the trajectory of his thinking as he encountered failures and setbacks with his computer programs. My attempted replication also failed, and that led me to try to unearth the five different programs he wrote, understand them, and figure out how one led to the next. They are not described in published papers, only in internal reports. My search for these reports led to their being made available on the Net.

What I found in my voyage were the early parts of a chain of thought that started with cybernetics, mathematics, and a plain-spoken computer; passed through "A City is Not a Tree"; paused to "make God appear in the middle of a field"; and ended with this fundamental design goal: *I try to make the volume of the building so that it carries in it all feeling. To reach this feeling, I try to make the building so that it carries my eternal sadness. It comes, as nearly as I can in a building, to the point of tears.*

# The Art, Science, and Engineering of Programming

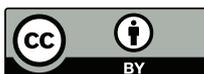

Notes on "Notes on the Synthesis of Form"

## 1 An Indian Village

"Christopher Alexander" is a household name for many computer scientists who have embraced his ideas surrounding patterns and pattern languages,[1] and some of them have even taken a shine to his more mystical ideas of the *quality without a name* and life / wholeness from "The Nature of Order" [10, 11, 13, 12].

The beginning of Christopher Alexander's career was "Notes on the Synthesis of Form" [5] (herein called "Notes"[2]) (1964), followed next by "A City is Not a Tree" [14] (1965). For those with technical or scientific backgrounds, what stood out in "Notes" was the mathematical approach, which sought to break a design problem into disjoint sets of requirements, presented as a tree. For computer scientists especially, this approach seemed to presage the modularity issues now called *cohesion* and *coupling*. The paper "A City is Not a Tree" seemed to some to come out of the blue: in it Alexander repudiated the idea of even trying to so definitely partition a complex design problem into disjoint, tree-like sets of interactions and concerns.[3] For me, who resisted reading "Notes" for decades, it was "Notes" that came out of the blue—it seemed not of the same sort as his other work, and thus it was a mystery how he came to his later ideas from these early ones. In most of his later work it was clear to me that he still wanted (or needed) a mathematical, scientific, objective, or rational foundation for his ideas, but that such a foundation escaped him, and so—I speculate—he came to rely on a spiritual or intuitive foundation. In a documentary of him done much later than "Notes," he said "we're trying to do something that no one else has ever tried to do in the 20th century…make God appear in the middle of a field" [25].

Here is a paradox: I suspect that in reviewing his own path and early work over against his later work, Alexander also considered "Notes" as coming out of the blue. In the preface to the paperback edition of "Notes" in 1971,[4] Alexander wrote this:

> *At the time I wrote this book, I was very much concerned with the formal definition of "independence," and the idea of using a mathematical method to discover systems of forces and diagrams which are independent.* But once the book was written, I discovered that it is quite unnecessary to use such a complicated and formal way of getting at the independent diagrams.

Also in 1971, he said this about the field called "Design Methods," partly spawned by "Notes": "…I would say forget it, forget the whole thing" [7].

---

[1] Since the publication of "Design Patterns" [22] in 1994, there have been hundreds of books published about design / software patterns, thousands of published pattern languages, and tens of thousands of published patterns.

[2] The book is derived from Alexander's Harvard Ph.D. dissertation, completed in 1962.

[3] To appreciate how strong that repudiation was, see Appendix E.

[4] The copy of "Notes" I read in 2015—the one that led to the investigation reported in this essay—was a sweet first edition I found at Powell's Bookstore in Portland Oregon. Had I instead read the 1971 edition with this preface.…

The story I uncovered of Alexander struggling to approach design formally is a story of trial and error, with error dominating. It's the story behind that preface. It's not a story of "if only": if only he were a better algorithmist or if only he made fewer mistakes, he might have succeeded in finding an algorithmic way to approach design. It's the story of the first few steps in the path Alexander took away from the formalisms of "Notes" toward the humanity of "A City is Not a Tree," and from there to "A Pattern Language" and to his most important contribution: teaching us that design requires human feeling.

**1.1 The Problem; Misfits; Homeostasis**

In "Notes" Christopher Alexander presents a detailed and extensive example problem: the redesign of a village in India of some 600 people to better suit present and future demands.[5] The essence of design as described in "Notes" is to minimize the number of *misfits*. One can think of this concept in the bluntest terms by considering a door that does not fit its door frame. This concrete notion is then taken in a metaphorical direction so that one can talk about the misfit of some aspect of a house to the lives of the people who live there. Alexander described it this way:

> *The same happens in house design. We should find it almost impossible to characterize a house which fits its context. Yet it is the easiest thing in the world to name the specific kinds of misfit which prevent good fit.[6] A kitchen which is hard to clean, no place to park my car, the child playing where it can be run down by someone else's car, rainwater coming in, overcrowding and lack of privacy, the eye-level grill which spits hot fat right into my eye, the gold plastic doorknob which deceives my expectations, and the front door I cannot find, are all misfits between the house and the lives and habits it is meant to fit. These misfits are the forces which must shape it, and there is no mistaking them. Because they are expressed in negative form they are specific, and tangible enough to talk about.*

Alexander analyses the problem of design by using the concept of a ***misfit variable***. A *misfit variable* is a binary variable that can take on the values 0 or 1. A value of 0 indicates that the condition represented by the variable *fits*. The value 1 indicates a misfit. This, though, is only part of the description of a design problem; the other part is how—or, more accurately, whether—misfit variables interact with each other. A connection between misfit variables means that any alteration of how one is treated (designed, constructed, positioned, etc.) can affect the other. Alexander presents an example to explain the basic idea of decomposition—a set of misfit variables represented as a network with links between nodes that interact:

---

[5] After completing his PhD and before publishing "Notes," Alexander was offered the opportunity to build such a village, but turned it down because (he said) he didn't know how to "harness the energy of the people in the village" to do the collaborative work of actually making a village [1].

[6] Alexander's later work on patterns, the quality without a name, centers, wholeness, and life is an attempt to codify the characteristics of "good fit."

**Notes on "Notes on the Synthesis of Form"**

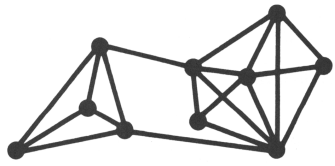

This is followed by this note:

> *Now, let us go back to the question of adaptation. Clearly these misfit variables, being interconnected, cannot adjust independently, one by one. On the other hand, since not all the variables are equally strongly connected (in other words there are not only dependences among the variables, but also* independences*), there will always be subsystems like those circled below, which can, in principle, operate fairly independently.*

…and this diagram:

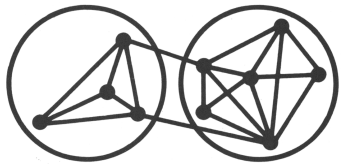

Seeing these diagrams, I immediately thought that the concepts in play are *cohesion* and *coupling*, concepts familiar in programming language and software circles as key to creating good modularity in code. Cohesion is the degree to which things that interact strongly with each other are kept together, and coupling is the degree to which things interact across boundaries. The idea is to partition software code into *modules*, each of which is a unit whose elements share information one way or another, typically information needed to co-design the elements in the unit. Information transfer across module boundaries should be minimal and through well-specified explicit protocols. To foreshadow an analogy I will use later, a module is like a tight-knit family where "there are no secrets." A family and its members can interact with another family, but in a formal, almost stranger-to-stranger manner.

But the work reported in "Notes" took place in the late 1950s, when these concepts were not known by these names, and the concepts themselves were perhaps only vaguely recognized. Instead of these concepts, Alexander appeals to *homeostasis*, a concept described and explored by W. Ross Ashby in "Design for a Brain" [18], first published in 1952.

Homeostasis is the tendency toward a relatively stable equilibrium between interdependent elements, and was one of the bases for thinking formally about living systems in the 1950s, a time that saw the birth of Artificial Intelligence. Ross Ashby was at the center of ideas in the circle that included cybernetics.

Alexander asks readers to imagine a set of lights that behave as follows:

1. if a light is on, every second there is a 50% probability it will turn off
2. if a light is off, every second there is a 50% probability it will turn on **if** it is connected to a light that is on

If all the lights happen to be off, there is no way for any of them to turn on again. Alexander explains that each light can be thought of as a misfit variable: the *off* state corresponds to fit; the *on* state corresponds to misfit. He continues as follows:

> *The fact that a light which is on has a 50–50 chance of going off every second, corresponds to the fact that whenever a misfit occurs efforts are made to correct it. The fact that lights which are off can be turned on again by connected lights, corresponds to the fact that even well-fitting aspects of a form can be unhinged by changes initiated to correct some other misfit because of connections between variables. The state of equilibrium, when all the lights are off, corresponds to perfect fit or adaptation. It is the equilibrium in which all the misfit variables take the value 0. Sooner or later the system of lights will always reach this equilibrium. The only question that remains is, how long will it take for this to happen? It is not hard to see that apart from chance this depends only on the pattern of interconnections between the lights.*

For example, if each light is connected to no others, then for each light, once it's off it stays off. If each light is connected to each of the others, then only when they all happen to (randomly) turn off will they all be off and stay off. This will eventually happen. In between these extremes, he argues, if there are clusters of well-connected lights with the clusters sparsely connected—or not connected at all—the time until they are all off will arrive sooner than if the lights are densely connected. Cohesion and coupling, we would say in the first quarter of the 21$^{st}$ century. But Alexander appears to start off with this question of the likelihood of a complex design problem consisting of misfit variables all turning to 0 simultaneously, and hence the strange (to me) nature of the mathematical argument in Appendix 2 in "Notes."

### 1.2 Alexander's Decomposition Problem

Alexander describes the algorithmic problem he endeavored to solve by using an example of partitioning a set $S_3$ into two subsets, $S_1$ and $S_2$. First, the misfits in $S_1$ should cohere somehow or "have something in common," as should the ones in $S_2$. He says that the easiest way to know that the misfits in a set cohere is that the interactions between them are dense, like this:[7]

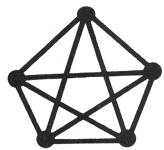

Second, because the realizations of $S_1$ and $S_2$ must be combined to form the realization of $S_3$ while taking into account the misfit links between them, there should be few links between $S_1$ and $S_2$.

---

[7] Looking at it from a homeostasis point of view, this is the worst choice because it takes about five times longer for this to settle down than a set of five isolated lights.

**Notes on "Notes on the Synthesis of Form"**

**1.3 Indian Village Problem Explained; Partitioning; Errors in the Data; HIDECS**

I was intrigued by the idea hinted at that a program written around 1960 could solve as complex a problem as the Indian Village redesign / rebuild. The essential problem was to take a set of design *requirements* (141 of them), a set of interactions among them (about 1400 of them), and partition them into groups that represent coherent design subtasks (more or less) or *components*. Alexander's approach was to create a *goodness* measure that would determine (numerically) how good a partition is. Then the idea was to *generate* disjoint partitions and *test* them using this measure—computer scientists call this algorithmic search technique "generate-and-test." The "Notes" Appendix included a pretty decomposition of the problem.

I tried to reproduce Alexander's results. I was immediately confused by the many clerical-like errors in the raw data supplied in the Appendix and the odd mathematical approach he took to creating his goodness measure. The clerical errors and sketchy definitions of terms made interpreting the apparently straightforward goodness measure difficult. Moreover, "Notes" did not contain a direct statement that the program hinted at actually produced the presented decomposition.

The references in "Notes" mention two research reports that seemed to promise explanations: I call them "HIDECS 2" and "HIDECS 3." I was unable to obtain them until long after the start of my investigation.

The full list of all the requirements is in Appendix A. Next comes a table of interactions between the requirements. Alexander wrote:

> *The links between these misfit variables are tabulated below. For the sake of simplicity, I allowed only one strength of link, so that $v = 1$, and for every pair of variables $v_{ij} = 0, 1,$ or $-1$. Further, the signs of the links are not indicated: as we shall see in Appendix 2* [of "Notes"]*, the decomposition turns out to be independent of the link signs. The table below simply shows those linked pairs of variables for which $v_{ij} = 1$ or $-1$.*

The table starts like this; the complete table is in Appendix B.

| | |
|---|---|
| 1 interacts with | 8, 9, 12, 13, 14, 21, 28, 29, 48, 61, 67, 68, 70, 77, 86, 101, 106, 113, 124, 140, 141 |
| 2 interacts with | 3, 4, 6, 26, 29, 32, 52, 71, 98, 102, 105, 123, 133 |
| 3 interacts with | 2, 12, 13, 17, 26, 76, 78, 79, 88, 101, 103, 119 |

The clerical-like errors occur in this list of interactions; there are 50 of them, each of the following form: "86 interacts with 3" is included but "3 interacts with 86" is excluded. All these erroneous 1-way interactions are shown in Appendix B.[8] Of the 50 errors in the interactions table, 30 involve requirement 33: "Fertile land to be used to best advantage."

---
[8] More about those errors is in Appendix F.

Richard P. Gabriel

The key to Alexander's mathematical analysis of complex decomposition problems and the goodness measure he creates is counting the number of links between sets of requirements. Before I had the source code for his program, these errors made it hard to understand his analysis and therefore his goodness measure.

In Appendix 1 of "Notes" Alexander presents the solution—or at least a solution—to the problem of decomposing the Indian Village design problem. Alexander wrote:

> *Analysis of the graph $G(M,L)$, shows us the decomposition pictured below, where M itself falls into four major subsets A, B, C, D, and where these sets themselves break into twelve minor subsets, A1, A2, A3, B1, B2, B3, B4, C1, C2, D1, D2, D3, thus:*

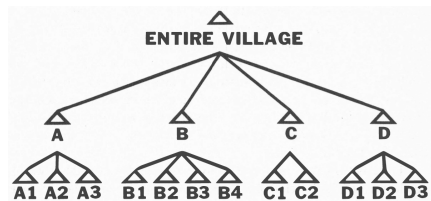

The breakdown is described like this:

| | |
|---|---|
| A1 contains requirements | 7, 53, 57, 59, 60, 72, 125, 126, 128 |
| A2 contains requirements | 31, 34, 36, 52, 54, 80, 94, 106, 136 |
| A3 contains requirements | 37, 38, 50, 55, 77, 91, 103 |
| B1 contains requirements | 39, 40, 41, 44, 51, 118, 127, 131, 138 |
| B2 contains requirements | 30, 35, 46, 47, 61, 97, 98 |
| B3 contains requirements | 18, 19, 22, 28, 33, 42, 43, 49, 69, 74, 107, 110 |
| B4 contains requirements | 32, 45, 48, 70, 71, 73, 75, 104, 105, 108, 109 |
| C1 contains requirements | 8, 10, 11, 14, 15, 58, 63, 64, 65, 66, 93, 95, 96, 99, 100, 112, 121, 130, 132, 133, 134, 139, 141 |
| C2 contains requirements | 5, 6, 20, 21, 24, 84, 89, 102, 111, 115, 116, 117, 120, 129, 135, 137, 140 |
| D1 contains requirements | 26, 29, 56, 67, 76, 85, 87, 90, 92, 122, 123, 124 |
| D2 contains requirements | 1, 9, 12, 13, 25, 27, 62, 68, 81, 86, 113, 114 |
| D3 contains requirements | 2, 3, 4, 16, 17, 23, 78, 79, 82, 83, 88, 101, 119 |

(Appendix C contains the requirements in A, B, C, D—the next level up.)

In addition to trying to decipher Alexander's approach, I tried several now-classical algorithms: K-Means clustering, Silhouette clustering, Karger's algorithm, and several of my own devising. For generate-and-test I used dynamic programming, greedy algorithms, simulated annealing, genetic programming, and some simple hill-climbing techniques. None came close to reproducing the above decomposition from "Notes."

The strong hint that Alexander used a program to come up with the shown decomposition—and a good hint, as it turns out—for what that program could be is the last few paragraphs of the last Appendix in "Notes":

**Notes on "Notes on the Synthesis of Form"**

> *Let us consider, lastly, the practical problem of finding that partition $\pi$, of the set M, for which this function $R(\pi)$[9] takes the smallest (algebraic) value.*
>
> *To find the best partition of a set S, we use a hill-climbing procedure which consists essentially of taking the partition into one-element subsets, computing the value of $R(\pi)$ for this partition, and then comparing with it all those partitions which can be obtained from it by combining two of its sets. Whichever of these partitions has the lowest value of $R(\pi)$ is then substituted for the original partition; and the procedure continues. It continues until it comes to a partition whose value of $R(\pi)$ is lower than that of any partition which can be obtained from it by combining two sets.[10]*
>
> *Another hill-climbing procedure, which finds a tree of partitions directly, goes in the opposite direction. It starts with the whole set S, and breaks it into its two most independent disjoint subsets, by computing $R(\pi)$ for a random two-way partition, and improving the partition by moving one variable at a time from side to side, until no further improvement is possible. It then repeats this process for each of the two subsets obtained, breaking each of them into two smaller subsets, and so on iteratively, until the entire set S is decomposed.[11]*
>
> *These and other methods have been programed for the IBM 7090, and are described in full elsewhere [3, 16]. It is important, and rather surprising, that the techniques do not suffer from the sampling difficulties often found in hill-climbing procedures, but gives extremely stable optima even for short computation times.*

**1.4 My Failed First Attempts to Recreate Alexander's Results**

My first thought was to try to recreate the twelve partitions Alexander showed. I had in front of me his goodness function and a variety of optimization algorithms; and I had been experimenting for a long time with simulated annealing, which is a probabilistic technique for approximating the global optimum of a given function. Here is Alexander's goodness function, $R(\pi)$:

$$R(\pi) = \frac{\frac{1}{2}m(m-1)\sum_{\pi}v_{ij} - l\sum_{\pi}S_\alpha S_\beta}{\left[(\sum_{\pi}S_\alpha S_\beta)(\frac{1}{2}m(m-1) - \sum_{\pi}S_\alpha S_\beta)\right]^{\frac{1}{2}}}$$

The tricky part for me was this:

$$\sum_{\pi} v_{ij}$$

and that because of the errors in the listed interactions. If there are $i, j$ where $v_{ij} = 0$ and $v_{ji} = 1$, then summing over $\pi$ needs to be interpreted to take that into account so that permutations of $\pi$ compute the same value for *R*—perhaps $v_{ij}$ and $v_{ji}$ need to be added in separately. That seemed messy. After a lot of exposition in the text, Alexander makes it (fairly) clear that it must be that $v_{ij} = v_{ji}$, and therefore, when we see $v_{ij} = 0$ and $v_{ji} = 1$, we can either throw out $v_{ji} = 1$ (that is, set $v_{ji} = 0$) or promote $v_{ij}$ (that is, set $v_{ij} = 1$). Because there was no way for me to know whether such an asymmetric interaction was real, my inclination was that it would be odd to have accidentally put 88 in the list of elements that interact with 3, where one really meant there was no interaction at all; therefore, I tended to promote interactions rather than delete them. That is, when 88 is in the interactions list for 3 but not vice versa, my program added 3 to the list of interactions for 88. It turns out I was wrong about that; the reason is in Appendix F.

---

[9] Alexander's goodness measure.
[10] This turns out to be the program BLDUP described in the HIDECS 3 report.
[11] This turns out to be the program described in the HIDECS 2 report.

However, I entertained that what Alexander intended was that $v_{ij}$ could take on the values 0, 1, 2, and so I adjusted and guessed at variants for $R(\pi)$. I will admit that the homeostatic approach to the mathematical treatment threw me for a loop—mostly because it seemed unnatural given the simple examples Alexander showed and what we know these days about cohesion, coupling, and clique detection. On top of that, the exposition in Appendix 2 of "Notes" was not the most rigorous I've seen, which meant there were places where I had to guess what he meant by some notation or mathematical move.

### 1.5 Finding the HIDECS Programs; Finding the Story

After many failed tries at reproducing the results in "Notes" I finally obtained the two HIDECS reports as well as a version of the HIDECS 2 program transliterated into Python [28]. At the same time, I obtained a paper entitled "The Determination of Components for an Indian Village" [4], in which Alexander shows a slightly different goodness measure from the one in "Notes."

The two HIDECS reports describe five different programs, each using a different approach to partitioning a design problem. The HIDECS 2 report describes one, and the HIDECS 3 report describes four, each a response to problems uncovered in previous programs. After receiving the new material I coded up (in Common Lisp) my own versions of most of them, but none of them produced exactly the decomposition in "Notes." However, that was not the interesting conclusion.

Until I found these reports, I was focussed on understanding Alexander's algorithmic approach by trying to retrace his footsteps—however blindly I was doing that. I believed that perhaps the clerical (and other?) errors alongside struggles with early 1960's computers, programming, and an immature algorithmic base were the reasons that his reported Indian Village decomposition did not match anything I could produce nor, as we'll see, anything his programs did produce. This focus continued until it finally dawned on me that I was observing Alexander coming to grips with the flaws of trying to use a formal, algorithmic approach to real-world design problems, that I was watching him transition from the "Notes" version of Alexander to the "A City is Not a Tree" version. And he was making this transition because the software was not

Notes on "Notes on the Synthesis of Form"

helping—because it could not help. The puzzles I encountered with my programs were the puzzles he struggled with.
    I didn't expect this.

### 1.6  Unfolding the HIDECS Programs; Studying Their Flaws

The program called HIDECS 2 was designed to separate components into clusters with minimal information transfer between them, meaning that the number of interaction links across cluster boundaries is small. Alexander was trying to solve the coupling part of the cohesion / coupling problem. Using the family analogy, he was trying to identify families in a population by finding clusters of people where people in each cluster don't do much with people in the other clusters.
    In the HIDECS reports Alexander calls the design requirements "vertices" or "misfit variables" and the interactions between them "links." HIDECS 2 proceeds by splitting the set of all the vertices into two disjoint subsets (partitions) using a random selection process that produces two subsets of, typically, unequal size. Next the program systematically tries moving single vertices from one subset to the other, one at a time, measuring the goodness of partition at each step, and selecting the best. This yields a binary partition of the set of vertices into disjoint subsets; the program moves ahead by doing the same process on the two partitions separately. The result is a binary tree: each node in the tree has exactly two subtrees below it. Computer scientists describe this strategy as a "top-down algorithm." Note also that the goodness measure needs to measure the goodness of a partition of only two sets.
    In my early investigations I had discovered that trying to find clusters by looking for weak coupling did not work well when the interactions were dense, as in the Indian Village problem. I also tried looking at cohesion as well as cohesion / coupling combined. In the main body of "Notes," Alexander shows what he calls "a typical graph" as part of his description of how to decompose design problems using his program: Figure 1a.
    Every program I wrote and every program in the HIDECS reports can decompose this. By way of contrast, a visualization of the network of interactions for the Indian Village problem is shown in Figure 1b. I sometimes wonder whether Alexander would have given up sooner on trying to devise decomposition methods to solve design problems had he been able to visualize the Indian Village network like this.
    Once one starts to look for strongly cohesive clusters instead of only loosely coupled ones in a dense network of interactions, overlap naturally occurs. I know that Alexander noticed this too. First, because playing with Alexander's earliest program and seeing it not do a good job or not doing a consistent job would lead anyone with curiosity to try alternatives. Second, because he said so in the HIDECS 3 report [16]:

> *HIDECS 2 has three important weaknesses:*
>     *1. The fact that the decomposition is made in a series of binary steps leads to certain 'mistakes,' since the holistic relatedness of system and subsystems is not properly taken into account.*

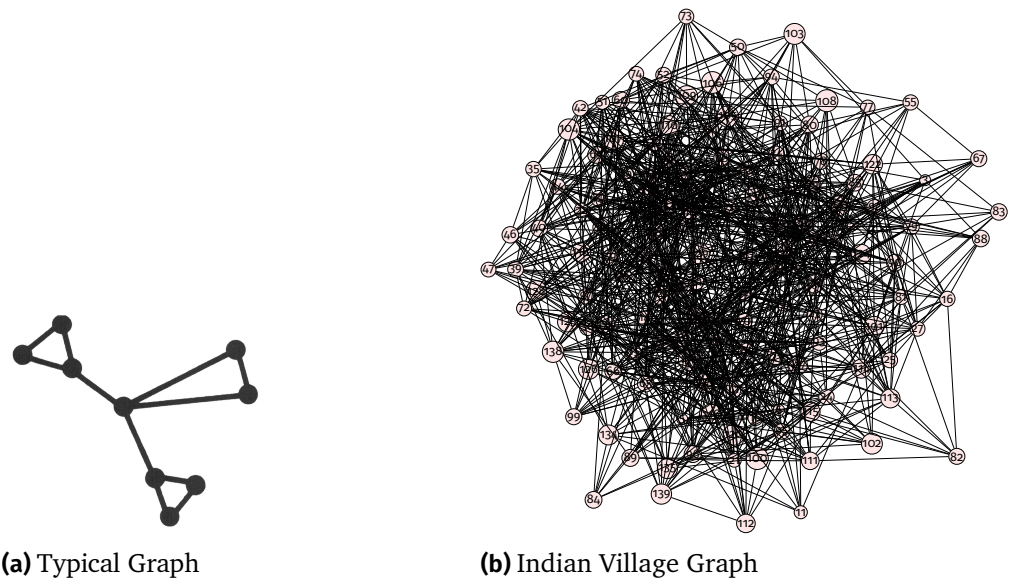

**(a)** Typical Graph  **(b)** Indian Village Graph

**Figure 1** Alexander's Typical Graph and Indian Village Actual Graph

> *2. The fact that the decomposition criterion INFO [the goodness measure] is based on very stringent assumptions about the nature of the system $G(M, L)$. Namely, that the elements of M are binary variables, that the two variable correlations are very small, and that the many variable correlations vanish altogether. These assumptions make it hard to find systems in the real world which the formalism of HIDECS 2 can adequately represent.*
>
> *3. The fact that the subsets of elements which make the most natural subsystems of a system are not always disjoint, but often overlap.*
>
> <div align="center">**(H3)**</div>

This was the passage that started me thinking that in pursuing Alexander's algorithmic journey, I had discovered a more important story: how did someone so wrapped up in a formal, mathematical approach to design turn his back on that and head in an intuitive, spiritual, artistic direction. I also started to believe that Alexander had only a partially formed idea about the design problem he was looking at, and that his struggles with his programs and their flaws was how he was uncovering what he was after. Nevertheless, I kept going with my examination of his struggles with software and my efforts to reproduce his results.

Once I had all the bits of source code I needed to understand what the HIDECS programs were doing, my interest in improving the results faded, as I suspect it did for Alexander. It became clear that the original program, HIDECS 2, being a randomized algorithm, could spit out a different partition each time it ran, but that there was a limit to how well they would measure out according to the goodness measure. Moreover, as far as I know, Alexander never reported a complete partitioning of the Indian Village problem. This makes it difficult to judge how well his original program did compared to my modern recoding[12] of his program running on modern hardware.[13] Alexander wrote in the HIDECS 2 report:

> ...*the program requires as input... LATIS, the number of starting sets for the hill-climbing algorithm to be chosen from the lattice.... The larger the value of LATIS selected, the more likely that the sampling procedure will discover the optimal TSET—but as the sample size increases, so does the amount of computer time used.*

My program running on my computer can support values for LATIS 50–500 times larger than his could for a given expected duration of computation.[14] For the goodness measure I decided to use the one he described in "Determination of Components," which is not quite the same as the one in the HIDECS 2 report, but it preserves ordering—if $G_D$ is the measure in "Determination of Components" and $G_H$ is the measure in HIDECS 2, then[15]

$$G_D(\pi_1) < G_D(\pi_2) \iff G_H(\pi_1) < G_H(\pi_2)$$

In general the results for the Indian Village were that his program found partitions with worse goodness measures than mine. The only directly stated example of a partition into exactly two sets is the partition of C into C1 and C2:

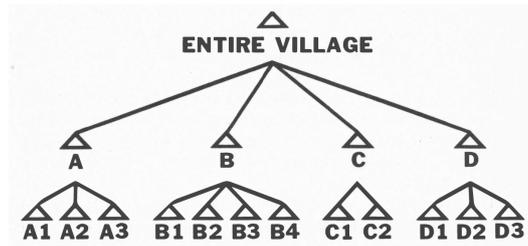

where

$$\text{C1} = 8, 10, 11, 14, 15, 58, 63, 64, 65, 66, 93, 95, 96, 99, 100, 112, 121, 130, 132, 133, 134, 139, 141 \quad (1)$$
$$\text{C2} = 5, 6, 20, 21, 24, 84, 89, 102, 111, 115, 116, 117, 120, 129, 135, 137, 140 \quad (2)$$

The goodness measures for Alexander's partition and the one my program produced using 250 times more starting sets is as follows, where smaller is better (−91 is better than −89):

|     | $G_D$  |
| --- | ------ |
| CA  | −89.60 |
| rpg | −91.60 |

---

[12] In Lispworks Common Lisp.
[13] On an Apple Mac Pro (2013): 3 GHz, 8-core, 64 GB RAM.
[14] This even though my programs use human-readable data structures and not Alexander's bit-level ones—I chose this approach to be able to understand what the programs are doing at the finest granularity.
[15] $G_H$ is HIDECS2-Actual and $G_D$ is HIDECS2-Decomp, as described below.

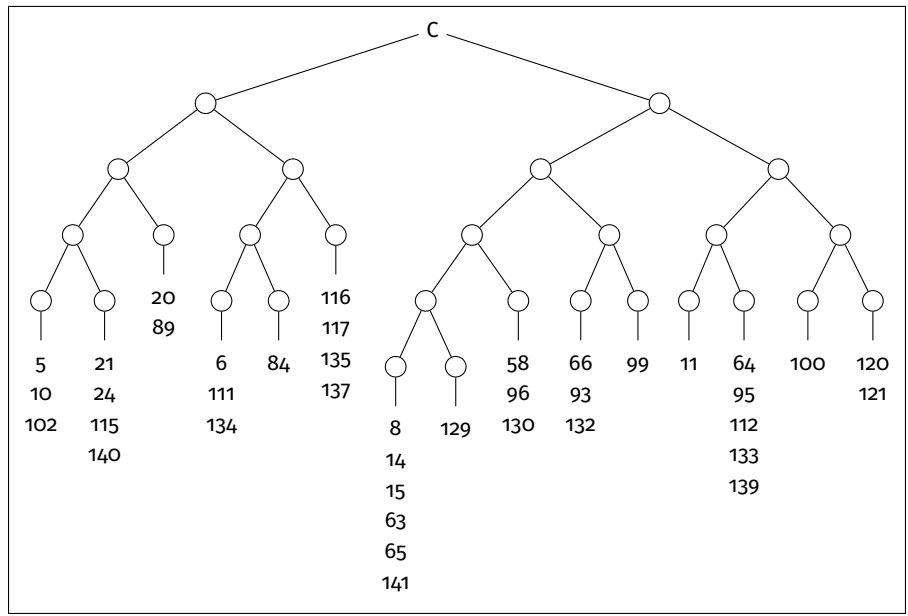

**Figure 2** Possible HIDECS 2 Decomposition for Alexander's Partition C

If HIDECS 2 actually produced this decomposition of C (into C1 and C2), it must have produced a full tree like the tree in Figure 2, and Alexander then hand-coalesced the lower nodes and leaves to get C1 and C2. Figure 2 actually shows the decomposition my program got, which as noted, measured at −91.60 for the first level (the equivalent of C1 and C2).

I ran my program to decompose C 100 times and got 70 different decompositions, each with the same measure for the top-level bifurcation (−91.60), with goodness measures at the leaf level ranging from −136.94 to −144.68. Figure 2 shows one of the (two) decompositions measuring at −144.68.

Before continuing, let's look at the various goodness functions Alexander describes.

## 1.7 The Goodness Measures

Understanding Alexander's explorations requires understanding both the algorithms and goodness measures he devised. The goodness measures and their intuitions are described in Appendix D.

| | |
|---|---|
| HIDECS2-Actual: | Defined in the HIDECS 2 Report. |
| HIDECS2-Decomp: | Defined in "The Determination of Components for an Indian Village" [4]. It seems to have been used for the "Notes" decomposition. |
| HIDECS2-Notes: | Defined in "Notes." |
| HIDECS2-rpg: | My cohesion / coupling goodness measure. |
| HIDECS3-BLDUP: | Defined in the HIDECS 3 Report and used for BLDUP. |
| HIDECS3-STABL: | Defined in the HIDECS 3 Report and used for STABL. |

**Notes on "Notes on the Synthesis of Form"**

**1.8 A Telltale Anomaly**

Alexander gave up on his top-down binary decomposition approach in HIDECS 3 because of an anomaly he observed (refer to Quote (H3). I discovered this anomaly while looking at the top two levels of decomposition in "Notes." I wanted to see how well Alexander's program did partitioning the Entire Village—the hardest partition of all—by trying to figure out what his first level down partition looked like. While doing that I found that the best partition at this top level did not produce the best partitions at the next level. That is, the top-down approach does not necessarily produce the best partition. The details of this exploration are in Appendix G.

**1.9 Examining Alexander's Elusive Results in "Notes"**

If some version of HIDECS 2 actually produced the decomposition in "Notes," it had to have produced a binary tree and not the one Alexander shows. I tried two different comparisons of Alexander's decomposition to ones my programs did: the first was to produce the decomposition down to level 4 (Alexander's lowest level) using HIDECS2-Actual, and the second using HIDECS2-rpg. The results were compared using HIDECS2-Notes; nothing interesting came of it: doing more thorough searches yielded slightly better decompositions (as measured), with HIDECS2-rpg producing better balanced partitions than HIDECS2-Actual. The results are in Appendix I.

**1.10 How Can We Figure Out Whether a Decomposition Is "Good"?**

One could argue that the relative goodnesses of two decompositions can be observed by looking at the texts of the requirements that are gathered together. Such a comparison would be a sort-of close reading. And such gathered-together texts can be the basis for an intuitive understanding of the decomposition. For the two decompositions above done by programs I wrote—*Entire Village (rpg$_1$)* (Figure 16) and *Entire Village (rpg$_2$)* (Figure 17)—I've coalesced the comparisons into a set of tables in Appendix J.

Another set of comparisons of the three decompositions is in Appendix K. For a modern algorithm for cohesion and coupling, see Appendix L.

## 2  A City Is Not a Tree

The HIDESC 2 report is firmly dedicated to breaking up a problem into a binary tree of disjoint sets of design concerns. The paper "A City is Not a Tree" repudiates that, and to many readers of Alexander's earliest work, that repudiation is unexpected.[16] However, the HIDECS 3 report shows Alexander slowly discovering that the programming ideas he pursued in HIDECS 2 were not going to yield a clean decomposition. See Quote (H3).

---

[16] Again, to appreciate how strong that repudiation was, see Appendix E.

Let's look at the programs in HIDECS 3 a bit. Each one addresses flaws in earlier programs.

The first flaw is that by going top down, HIDECS 2 never looks at the total, fine-grained partition presented by the leaves of the binary tree. The approach in the first HIDECS 3 program (BLDUP) is to start with a partition of the vertices into sets of single elements—for the Indian Village problem, this is 141 sets. The program systematically tries combining pairs of partitions, measuring the goodness of the entire partition; to do this, Alexander extended the HIDECS 2 measure (HIDECS2-Actual) to HIDECS3-BLDUP (Appendix D), which can operate on multiple sets in a partition. Alexander justifies this as follows:

> *However, the defect of any algorithm which partitions M into two subsets at a time, is that it does not pay sufficient attention to the gestalt, or overall pattern of the subsystem, and therefore introduces a bias which by any reasonable intuition is a 'mistake.'... In BLDUP, the decomposition criterion, though still essentially the same as that used in HIDECS 2, has been extended so that not only 2-way, but 3-way, 4-way, etc. partitions can all be compared with one another. This means that the decomposition into subsystems need not be defined stepwise, but can be defined all at once, and the holistic nature of the system thereby better preserved.*

Alexander's elegant demonstration of this is in Appendix H.

The program BLDUP produces a decomposition into disjoint sets, not a tree. Alexander observed that BLDUP's crude contraction approach combined with a coupling-based goodness measure which varied only very slowly with changing decompositions didn't work very well, so BLDUP was immediately discarded. Alexander quickly moved on to STABL, which uses a different approach and a new cohesion measure. Importantly, Alexander also wrote about the experimental setup:

> *In STABL, SIMPX, and EQCLA, the elements of the system are no longer assumed to be binary variables, indeed variables at all. The elements of M may be elements of any kind, and the links between elements, though still only permissible between two elements at a time, may be of any sort whatever. In all three cases, as in BLDUP, the subsystems are defined simultaneously, not sequentially.*

This means that the intuition he had earlier that looking at the design problem as a problem of homeostasis was not serving him well.[17]

The second program (STABL) proceeds by starting with a partition into single-vertex sets and then systematically tries to move one vertex at a time from the set it happens to be in to each of the other sets, one at a time. Alexander also created a new goodness measure that looks at both coupling and cohesion (HIDECS3-STABL, described in Appendix D)—that is, it looks at both how strongly each vertex is linked to other vertices in the same potential partition as well as how much information is transmitted from one partition to others. The algorithms using the earlier goodness measures try to minimize those measures—that is, minimize coupling; this algorithm tries to maximize the goodness measure—that is, maximize cohesion while minimizing coupling. Keep in mind he likely did not have available the named concepts of cohesion and coupling.

---

[17] Does that also mean that improving the speed of designing and building through decomposition is in jeopardy?

Alexander makes some interesting remarks. Talking about HIDECS3-STABL, which he calls "EXP," he wrote:

> *This function EXP varies sharply even over slight variation in the decomposition. The crude hill-climb by successive contractions, used in BLDUP, is therefore unsuitable for STABL. In fact, in experience, even for small and simple systems, a hill-climb based on contraction failed to find the decomposition with the best value of EXP. Instead STABL is based on the following procedure....*

In a footnote on the same page, Alexander wrote:

> *Actually EXP differs slightly from the criterion function given in (1963 b);[18] the changes make it more continuous in the search space; the original function had such severe discontinuities that the hill climb would not work at all.*

EXP (that is, HIDECS3-STABL) indeed seems to compute large values. Alexander goes on to describe the program, STABL, and in it he inadvertently touches on a flaw he seems to have missed:

> *Start with the unit decomposition in SETS, as before. The basic operation consists of moving one element, out of the set it happens to be in, and adding it to each of the other sets in turn, computing EXP for each new decomposition so obtained. This is done for each element. The best of all the decompositions so obtained is thus the best decomposition to be obtained by moving a single element. The program makes this change; and then begins the cycle again, The program terminates when it finds a decomposition whose value of EXP is higher than that for any decomposition obtainable from it by moving a single element.*

To understand the flaw, we need to look at a new example Alexander introduces; see Figure 3. Alexander wrote that his program produces the decomposition shown in Figure 4a. This measures as $36,862.235$ using HIDECS3-STABL. However, my recoding / interpretation comes out with the decomposition shown in Figure 4b as the best, measuring as $265,361.889$ (larger is better).

---

[18] This reference—(1963 b)—is to a paper by Alexander titled "The Most Stable Decomposition of a System into Subsystems" which was submitted in 1963 to the journal *Information and Control* published by Academic Press (now renamed *Information and Computation* published by Elsevier), but perhaps it was never published. I could not find it in any volume of that journal from 1961 through 1965 nor anywhere else.

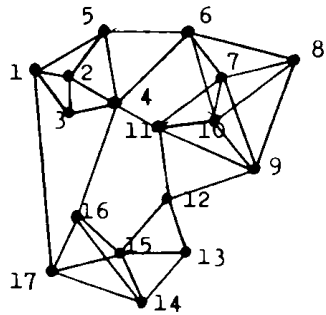

**■ Figure 3** HIDECS3-Graph

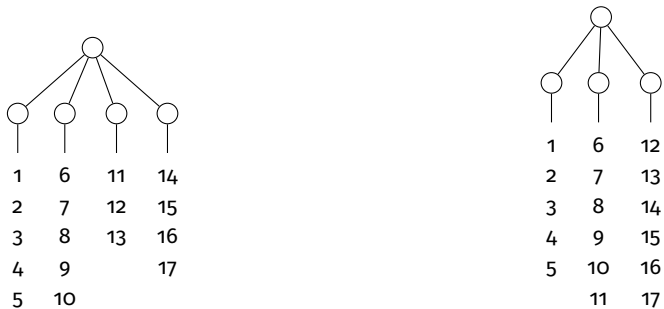

**(a)** Alexander's Decomposition  **(b)** rpg's Decomposition

**■ Figure 4** HIDECS3-Graph Decompositions

The issue is that when Alexander says "the best of all the decompositions so obtained is thus the best decomposition to be obtained by moving a single element," he overlooks the possibility that there could be ties, and thus he either has to make a choice and live with it moving forward, or do a tree-search—not simply a hill-climbing search. For this problem, the first pass through yields 74 possible first moves, all with the same goodness measure. For a run my program did, there were subsequent choice points with 2, 42, 3, 2, 20, 2, and 2 available choices.

I found no evidence that he was aware of this. Note that my program does an exhaustive search to find the best decomposition (according to HIDECS3-STABL).

**2.1 Three of My Puzzles Answered; One Puzzle Raised**

At this point, three of my puzzles have been handled. First, the goodness measure in "Notes" seems not to have been used directly to create the Indian Village decomposition shown there. That is, HIDECS 2 talks about using HIDECS2-Actual, but the Indian Village decomposition was created using HIDECS2-Decomp—as revealed in "The Determination of Components for an Indian Village" [4]. Neither of these goodness measures is HIDECS2-Notes, though when used on partitions into two sets, HIDECS2-Decomp and HIDECS2-Notes compute the same value.

Second, although Alexander in "Notes" introduces the bottom-up approach first—essentially BLDUP—it is the top-down approach of breaking down the original prob-lem into binary partitions that created the shown decomposition. Early on I was shunted away from the top-down approach because the partitions shown were not binary.

Third, the strange treatment of the "misfit variables" as binary (random) variables has been discarded, which better corresponded with my more modern-day view of the problem as a cohesion / coupling situation.

The new puzzle: If the Indian Village decomposition in "Notes" is supposed to be a good one—one deserving to be shown, dissected, and discussed at length—why did Alexander write in the HIDECS 3 report that "HIDECS 2 has three important weaknesses" (refer to Quote (H3)) and then go on to investigate four programs to correct those flaws?

## 2.2 Alexander Abandons Decomposition into Disjoint Sets

The last two of the four programs in the HIDECS 3 report are based on giving up on creating a partition into disjoint sets. Alexander wrote:

> *Finally, in SIMPX and EQCLA, the subsystems are defined in such a way that they overlap instead of being disjoint. In fact, in these two cases the decomposition, instead of being a tree, is a lattice.*

Once the first move was made to working with cohesion, Alexander moved more strongly in that direction. In 1957 a pair of researchers came up with an improvement to one of the first clique-detection algorithms: they were Frank Harary and Ian Ross [24]; Alexander adopted this algorithm for SIMPX (by direct reference to their paper). The essential idea is that a partition is very strong when each vertex interacts with every other one—this is the definition of a clique. For example, if there are three vertices, each interacts with the other two; if there are four, each interacts with the other three—these are examples of the most cohesive cliques. The third and fourth of the programs in HIDECS 3 are variations on this. In his typical graph (Figure 1a), one can see three strongly interacting triangles of vertices; these are cliques.

Alexander noticed such tight cohesions in the HIDECS 2 paper and program. While partitioning a set into subsets, when the program notices such complete graphs, it does not try to subdivide them.

The Harary-Ross algorithm has flaws, as reported by Harary in his 1969 text, "Graph Theory" [23]. Instead of using that algorithm, I used a more modern one, the Tomita variant of the Bron-Kerbosch algorithm. In 1967, Edward Bierstone and Allen Bernholtz developed a semilattice recomposition program described in their report "HIDECS-RECOMP PROCEDURE" [19]. I coded it up to visualize the produced decompositions. Rather than trying to describe the algorithm, I'll quote Bierstone and Bernholtz:

> *By formulating the system of minute requirements and interactions for the design problem, and employing a mathematical procedure to decompose the system into subsystems and hierarchically recombine these subsystems, we allow the specific*

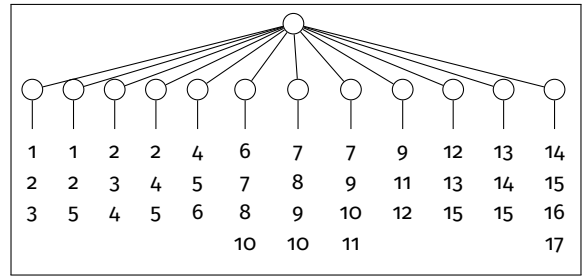

**Figure 5** Tomita Decomposition for HIDECS3-Graph

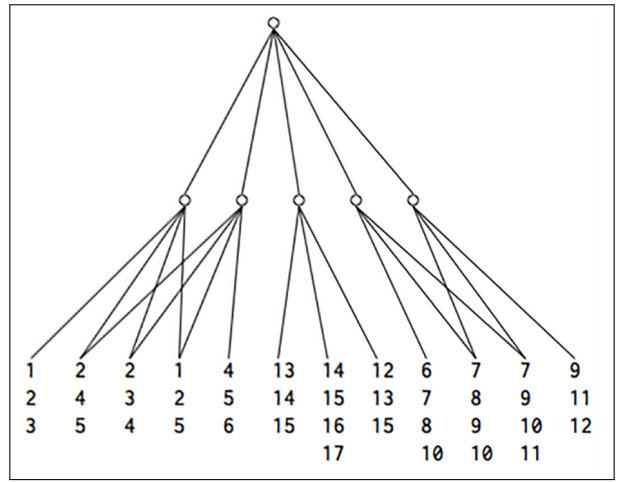

**Figure 6** Bierstone-Bernholtz Recomposition of the Tomita Decomposition for HIDECS3-Graph

> *problem to generate its own structure based on internal interaction of requirements, instead of arbitrarily dividing the problem into acoustics, zoning, circulation, and so on, or of starting with form preconceptions like "bedroom." The system of subsystems, furthermore, is structured as a semi-lattice (that is, overlap is possible at all levels of the hierarchy), so that we avoid the natural tendency to divide the problem into an artificial tree structure, in which subproblems may be combined hierarchically but do not overlap on any level. The semi-lattice structure of the problem is based only on causal interactions between elements, so that form decisions are not made until after the problem is structured.*

Essentially, the algorithm looks at a partition into sets, computes a reasonable set of interactions, and creates a semilattice structure that represents those interactions. It's a kind of complexity visualization. The decomposition Tomita comes up with for HIDECS3-Graph is shown in Figure 5. The recomposition Bierstone-Bernholtz comes up with is shown in Figure 6; it mainly shows the overlap complexity.

A final example of a design problem Alexander looked at is in Appendix P.

**Notes on "Notes on the Synthesis of Form"**

## <span>3</span>  Slowly Dawning Insights

During my investigations I was struck by the cold abstractness of the problem statement: 141 vertices and ∼1400 links binding them together. However, these requirements came from real people and state real issues. Alexander wrote:

> *All these misfit variables are stated here in their positive form; that is, as needs or requirements which must be satisfied positively in a properly functioning village. They are, however, all derived from statements about potential misfits: each one represents some aspect of the village which could go wrong, and is therefore a misfit variable....*

Moreover, the vertices are broken into 13 groups: *Religion and Caste*; *Social Forces*; *Agriculture*; *Animal Husbandry*; *Employment*; *Water*; *Material Welfare*; *Transportation*; *Forests and Soils*; *Education*; *Health*; *Implementation*; *Regional, Political, and National Development*; here is a selection from each group—the complete set is in Appendix A:

- 7. Cattle treated as sacred, and vegetarian attitude.
- 12. Men's groups chatting, smoking, even late at night.
- 36. Protection of crops from thieves, cattle, goats, monkeys, etc.
- 53. Upgrading of cattle.
- 65. Diversification of villages' economic base—not all occupations agricultural.
- 67. Drinking water to be good, sweet.
- 79. Provision of cool breeze.
- 98. Daily produce requires cheap and constant (monsoon) access to market.
- 106. Young trees need protection from goats, etc.
- 112. Access to a secondary school.
- 125. Prevent malnutrition.
- 129. Factions refuse to cooperate or agree.
- 133. Social integration with neighboring villages.

In "Notes" Alexander wrote:

> *Above all, the designer must resist the temptation to summarize the contents of the tree in terms of well-known verbal concepts. He must not expect to be able to see for every* [set] *S* [in a partition of the tree] *some verbal paradigm like "This one deals with the acoustic aspects of the form." If he tries to do that, he denies the whole purpose of the analysis, by allowing verbal preconceptions to interfere with the pattern which the program shows him. The effect of the design program is that each set of requirements draws his attention to just one major physical and functional issue, rather than to some verbal or preconceived issue. It thereby forces him to consolidate the physical ideas present in his mind as seedlings, and to make physical order out of them.*

While trying to reproduce the decomposition in "Notes," I entertained the hypothesis that Alexander made it by hand, and that he looked at the realities expressed in the requirement statements. Some of my speculative, pre-HIDECS-informed programs took into account the 13 groups or various other groupings of them based on what they meant. And in fact, when Alexander describes his decomposition, he spins a story of how they are connected. Here is one such, starting with an overview of the four main partitions:

> *The four main diagrams are roughly these: A deals with cattle, bullock carts, and fuel; B deals with agricultural production, irrigation, and distribution; C deals with the communal life of the village, both social and industrial; D deals with the private life of the villagers, their shelter, and small-scale activities. Of the four, B is the largest, being of the order of a mile across, while A, C, D, are all more compact, and fit together in an area of the order of 200 yards across.*

It's important to notice that this description implies that in the Indian Village design problem, it's possible for two design concerns to interact simply because they are near each other in the real world. He goes on to describe A1—he starts with a diagram:

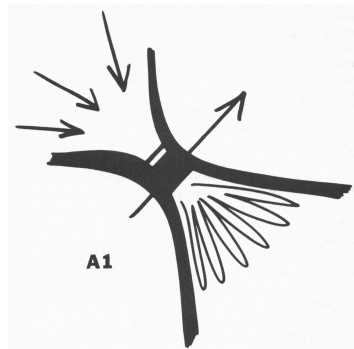

> *The sacredness of cattle (7) tends to make people unwilling to control them, so they wander everywhere eating and destroying crops, unless they are carefully controlled. Similarly, the need to upgrade cattle (53) calls for a control which keeps cows out of contact with roaming scrub bulls; and further calls for some sort of center where a pedigree bull might be kept (even if only for visits); and a center where scrub bulls can be castrated. Cattle diseases (57) are mainly transferred from foot to foot, through the dirt—this can be prevented if the cattle regularly pass through a hoof bath of disinfecting permanganate. If milk (59) is to be sold cooperatively, provision must be made for central milking (besides processing); if cows are milked at home, and the milk then pooled, individual farmers will adulterate the milk.[19] Famine prevention (72), the prevention of malnutrition (125), and price assurance for crops (128) all suggest some kind of center offering both storage, and production of nourishing foods (milk, eggs, groundnuts). If the village-level worker (126)*

---

[19] This statement is perhaps an opportunity to explore how much Alexander's knowledge of the real Indian Village influenced his partition. He seems compelled to note that "individual farmers will adulterate the milk," but none of the requirements in A1 appear to have anything to do with that. In fact, none of the whole set of requirements do.

*is to come often to the village and help, quarters must be provided for him here. Animal traction (60) calls for access to and from the cattle stalls (A2) on the one hand, and the road on the other.*

## 4 The Search for Beauty

Alexander's post-"Notes" career was not about finding better methods to decompose design problems, but about understanding what makes wonderful things and places wonderful. I call this wonderfulness *beauty*; Alexander avoided that word. We begin to see this search in "A City is Not a Tree" (my emphases added):

*It is vital that we discover the property of old towns which gave them* life.... *Too many designers today seem to be yearning for the physical and plastic characteristics of the past, instead of searching for the abstract* ordering principle *which the towns of the past happened to have, and which our modern conceptions of the city have not yet found. These designers fail to put new* life *into the city,... they fail to unearth its* inner nature.

This inner nature starts with semilattice structure and rich human interactions such structures support—it has more to do with human well-being than with artistic beauty. "A Pattern Language" [15] and "The Timeless Way of Building" [8] are about what makes cities, towns, and buildings have *life* or, to use Alexander's phrase the *quality without a name* (QWAN). Alexander and his team formulated the concept of *pattern* and catalogued 253 instances of spatial configurations that supported well-being in everyday, ordinary life. For example, a narrow ledge on the side of a building feels precarious, and most people only put out laundry or store bicycles there. A SIX-FOOT BALCONY (pattern #167), however, affords a sense of ease and safety, room for a small table and chairs, an invitation to connect life inside the building with life on the street, integrating the urban fabric both socially and physically. In "Timeless," Alexander wrote of QWAN:

*There is a central quality which is the root criterion of life and spirit in a man, a town, a building, or a wilderness. This quality is objective and precise, but it cannot be named,... in order to define this quality in buildings and in towns, we must begin by understanding that every place is given its character by certain patterns of events that keep on happening there. These patterns of events are always interlocked with certain geometric patterns in the space. Indeed, as we shall see, each building and each town is ultimately made out of these patterns in the space, and out of nothing else: they are the atoms and the molecules from which a building or a town is made.*

Alexander came up with the Indian Village requirements while living in a village in India and observing the complex, very human nature of that village. Unable to fathom it, he embarked on his search for automated decomposition, full of hope. In the end the complexity of Figure 1b defeated that search and taught him the need for rich structure, followed by QWAN. In "Timeless" Alexander continued:

> *What happens in a world—a building or a town—in which the patterns have the quality without a name, and are alive?*
>
> *...every part of it, at every level, becomes unique. The patterns which control a portion of the world, are themselves fairly simple. But when they interact, they create slightly different overall configurations at every place. This happens because no two places on earth are perfectly alike in their conditions. And each small difference, itself contributes to the difference in conditions which the other patterns face....*
>
> *Nature is never modular. Nature is full of almost similar units (waves, raindrops, blades of grass)—but though the units of one kind are all alike in their broad structure, no two are ever alike in detail.*

Alexander didn't stop at QWAN and patterns; he noticed that there was something more or different going on that makes, for example, one Turkish prayer carpet "better" than another, have more *life*, be more *whole*; it's what separates poetry from prose. Alexander lays out the idea of an underlying order of, well, everything in "The Nature of Order" [10]:

> *What we call "life" is a general condition which exists, to some degree or other, in every part of space: brick, stone, grass, river, painting, building, daffodil, human being, forest, city. And further: The key to this idea is that every part of space—every connected region of space, small or large—has some degree of life, and that this degree of life is well-defined, objectively existing, and measurable.*

Order—be it the geometry in natural systems or good artifacts—can be organized around fifteen fundamental geometric properties of what Alexander calls "centers," design becomes *unfolding*....

But that's another essay...or book.

## 5   *The I, That Blazing One*

What can we learn from these investigations? Christopher Alexander's journey was of slowly dawning insights not a grand "Aha!" First are the small insights, insights about the problems of decomposition, cohesion, and coupling viewed during the early days of computing.

Although he did not have the concepts of cohesion and coupling as they are now known, he navigated the waters between them. He was not shy about using techniques and algorithms invented by others: some randomized algorithms already existed and were generally known in the late 1950s; clique detection algorithms were known and Alexander acknowledges using one. Alexander and Manheim were not inept programmers—the HIDECS programs were written in assembly language and exhibited a sophisticated use of so-called "bumming" techniques.[20]

---

[20] Bum: "to make highly efficient, either in time or space, often at the expense of clarity."

**Notes on "Notes on the Synthesis of Form"**

Next is that Alexander was using the software he was creating to teach him about the problem he set out to solve—his understanding of the problem improved as the flaws in his programs revealed themselves; sometimes he tried to improve the programs, and other times he reformulated the problem.

Finally—and most importantly—Alexander's struggles taught him to look away from formalism—to look elsewhere—to understand design. Note the progression of thought from these very early investigations to those near the end of his career. This is his big insight: design requires human feeling. Imagine the mind that progressed as follows, starting with "Notes":

> *The tree of sets this decomposition gives is, within the terms of this book, a complete structural description of the design problem defined by M; and it therefore serves as a program for the synthesis of a form which solves this problem. . . .*
>
> *The organization of any complex physical object is hierarchical. It is true that, if we wish, we may dismiss this observation as an hallucination caused by the way the human brain, being disposed to see in terms of articulations and hierarchies, perceives the world. On the whole, though, there are good reasons to believe in the hierarchical subdivision of the world as an objective feature of reality.* [5]

That was originally from the early 1960s; next from "A City is Not a Tree":

> *For the human mind, the tree is the easiest vehicle for complex thoughts. But the city is not, cannot and must not be a tree. The city is a receptacle for life. If the receptacle severs the overlap of the strands of life within it, because it is a tree, it will be like a bowl full of razor blades on edge, ready to cut up whatever is entrusted to it. In such a receptacle life will be cut to pieces. If we make cities which are trees, they will cut our life within to pieces.* [14]

Around the same time, in another essay:

> *Myself, as some of you know, originally a mathematician, I spent several years, in the early sixties, trying to define a view of design, allied with science, in which values were also let in by the back door. I too played with operations research, linear programming, all the fascinating toys, which mathematics and science have to offer us, and tried to see how these things can give us a new view of design, what to design, and how to design.*
>
> *Finally, however, I recognized that this view is essentially not productive, and that for mathematical and scientific reasons, if you like, it was essential to find a theory in which value and fact are one, in which we recognize that here is a central value, approachable through feeling, and approachable by loss of self, which is deeply connected to facts, and forms a single indivisible world picture, within which productive results can be obtained.* [6]

Alexander was talking about what was called in the 1960s the "Design Methods Movement," of which he was considered a pioneer. In 1971 he was quoted as saying:

> *I've disassociated myself from the field.... There is so little in what is called "design methods" that has anything useful to say about how to design buildings that I never even read the literature anymore.... I would say forget it, forget the whole thing....* [7]

Then in "The Nature of Order, Book 4":

> *The I, that blazing one, is something which I reach only to the extent that I experience, and make manifest, my feeling. What feeling, exactly? What exactly am I aiming for in a building, in a column, in a room? How do I define it for myself, so that I can feel it clearly, so that it stands as a beacon to steer me in what I do every day?...*
>
> *What I aim for is, most concretely, sadness. I try to make the volume of the building so that it carries in it all feeling. To reach this feeling, I try to make the building so that it carries my eternal sadness. It comes, as nearly as I can in a building, to the point of tears.* [12]

We see in his early mind what his mind became. When we read the backstories in the HIDECS reports and read carefully the words in his formal publications, we learn that the reality of the computer and the poverty of programming languages were stern teachers, teaching Alexander that cold abstraction requires a warm human hand and experienced (tear-filled) eyes, that machines can be partners for exploration, and that a city is not a tree.

## A   Requirements

Quoting Christopher Alexander in "Notes":

> *Here is a worked example, taken from a recent paper, "The Determination of Components for an Indian Village"* [4]. *The problem treated is this. An agricultural village of six hundred people is to be reorganized to make it fit present and future conditions developing in rural India.*

These are the design considerations used in the example.

**Religion and Caste**
1. Harijans regarded as ritually impure, untouchable, etc.
2. Proper disposal of dead.
3. Rules about house door not facing south.
4. Certain water and certain trees are thought of as sacred.
5. Provision for festivals and religious meetings.
6. Wish for temples.
7. Cattle treated as sacred, and vegetarian attitude.

**Notes on "Notes on the Synthesis of Form"**

8. Members of castes maintain their caste profession as far as possible.
9. Members of one caste like to be together and separate from others, and will not eat or drink together.
10. Need for elaborate weddings.

**Social Forces**
11. Marriage is to person from another village.
12. Extended family is in one house.
13. Family solidarity and neighborliness even after separation.
14. Economic integration of village on payment-in-kind basis.
15. Modern move toward payment in cash.
16. Women gossip extensively while bathing, fetching water, on way to field latrines, etc.
17. Village has fixed men's social groups.
18. Need to divide land among sons of successive generations.
19. People want to own land personally.
20. People of different factions prefer to have no contact.
21. Eradication of untouchability.
22. Abolition of Zamindari and uneven land distribution.
23. Men's groups chatting, smoking, even late at night.
24. Place for village events—dancing, plays, singing, etc., wrestling.
25. Assistance for physically handicapped, aged, widows.
26. Sentimental system: wish not to destroy old way of life; love of present habits governing bathing, food, etc.
27. Family is authoritarian.
28. Proper boundaries of ownership and maintenance responsibility.
29. Provision for daily bath, segregated by sex, caste, and age.

**Agriculture**
30. Efficient and rapid distribution of seeds, fertilizer, etc., from block HQ.
31. Efficient distribution of fertilizer, manure, seed, from village storage to fields.
32. Reclamation and use of uncultivated land.
33. Fertile land to be used to best advantage.
34. Full collection of natural manure (animal and human).
35. Protection of crops from insects, weeds, disease.
36. Protection of crops from thieves, cattle, goats, monkeys, etc.
37. Provision of storage for distributing and marketing crops.
38. Provision of threshing floor and its protection from marauders.
39. Best cotton and cash crop.
40. Best food grain crop.
41. Good vegetable crop.

42. Efficient plowing, weeding, harvesting, leveling.
43. Consolidation of land.
44. Crops must be brought home from fields.
45. Development of horticulture.
46. Respect for traditional agricultural practices.
47. Need for new implements when old ones are damaged, etc.
48. Scarcity of land.
49. Cooperative farming.

**Animal Husbandry**
50. Protected storage of fodder.
51. Improve quality of fodder available.
52. Improve quantity of fodder available.
53. Upgrading of cattle.
54. Provision for feeding cattle.
55. Cattle access to water.
56. Sheltered accommodation for cattle (sleeping, milking, feeding).
57. Protection of cattle from disease.
58. Development of other animal industry.
59. Efficient use and marketing of dairy products.
60. Minimize the use of animal traction to take pressure off shortage.

**Employment**
61. Sufficient fluid employment for laborers temporarily (seasonally) out of work.
62. Provision of cottage industry and artisan workshops and training.
63. Development of village industry.
64. Simplify the mobility of labor, to and from villages, and to and from fields and industries and houses.
65. Diversification of villages' economic base—not all occupations agricultural.
66. Efficient provision and use of power.

**Water**
67. Drinking water to be good, sweet.
68. Easy access to drinking water.
69. Fullest possible irrigation benefit derived from available water.
70. Full collection of underground water for irrigation.
71. Full collection of monsoon water for use.
72. Prevent famine if monsoon fails.
73. Conservation of water resources for future.
74. Maintenance of irrigation facilities.
75. Drainage of land to prevent waterlogging, etc.
76. Flood control to protect houses, roads, etc.

**Notes on "Notes on the Synthesis of Form"**

**Material Welfare**

77. Village and individual houses must be protected from fire.
78. Shade for sitting and walking.
79. Provision of cool breeze.
80. Security for cattle.
81. Security for women and children.
82. Provision for children to play (under supervision).
83. In summer people sleep in open.
84. Accommodation for panchayat records, meetings, etc.
85. Everyone's accommodation for sitting and sleeping should be protected from rain.
86. No overcrowding.
87. Safe storage of goods.
88. Place to wash and dry clothes.
89. Provision of goods, for sale.
90. Better provision for preparing meals.
91. Provision and storage of fuel.
92. House has to be cleaned, washed, drained.
93. Lighting.

**Transportation**

94. Provision for animal traffic.
95. Access to bus as near as possible.
96. Access to railway station.
97. Minimize transportation costs for bulk produce (grain, potatoes, etc.).
98. Daily produce requires cheap and constant (monsoon) access to market.
99. Industry requires strong transportation support.
100. Provision for bicycle age in every village by 1965.
101. Pedestrian traffic within village.
102. Accommodation for processions.
103. Bullock cart access to house for bulk of grain, fodder.

**Forests and Soils**

104. Plant ecology to be kept healthy.
105. Insufficient forest land.
106. Young trees need protection from goats, etc.
107. Soil conservation.
108. Road and dwelling erosion.
109. Reclamation of eroded land, gullies, etc.
110. Prevent land erosion.

**Education**
111. Provision for primary education.
112. Access to a secondary school.
113. Good attendance in school.
114. Development of women's independent activities.
115. Opportunity for youth activities.
116. Improvement of adult literacy.
117. Spread of information about birth control, disease, etc.
118. Demonstration projects which spread by example.
119. Efficient use of school; no distraction of students.

**Health**
120. Curative measures for disease available to villagers.
121. Facilities for birth, pre- and post-natal care, birth control.
122. Disposal of human excreta.
123. Prevent breeding germs and disease starters.
124. Prevent spread of human disease by carriers, infection, contagion.
125. Prevent malnutrition.

**Implementation**
126. Close contact with village-level worker.
127. Contact with block development officer and extension officers.
128. Price assurance for crops.
129. Factions refuse to cooperate or agree.
130. Need for increased incentives and aspirations.
131. Panchayat must have more power and respect.
132. Need to develop projects which benefit from government subsidies.

**Regional, Political, and National Development**
133. Social integration with neighboring villages.
134. Wish to keep up with achievements of neighboring villages.
135. Spread of official information about taxes, elections, etc.
136. Accommodation of wandering caste groups, incoming labor, etc.
137. Radio communication.
138. Achieve economic independence so as not to strain national transportation and resources.
139. Proper connection with bridges, roads, hospitals, schools.
140. Develop rural community spirit: destroy selfishness, isolationism.
141. Prevent migration of young people and harijans to cities.

**Notes on "Notes on the Synthesis of Form"**

## B  Interactions

The requirements for the Indian Village problem shown in Appendix A are linked; each link represents a design-related interaction. These are the design interactions used in the example.

| | |
|---|---|
| 1 interacts with | 8, 9, 12, 13, 14, 21, 28, 29, 48, 61, 67, 68, 70, 77, 86, 101, 106, 113, 124, 140, 141 |
| 2 interacts with | 3, 4, 6, 26, 29, 32, 52, 71, 98, 102, 105, 123, 133 |
| 3 interacts with | 2, 12, 13, 17, 26, 76, 78, 79, 88, 101, 103, 119 |
| 4 interacts with | 2, 5, 6, 17, 29, 32, 45, 56, 63, 71, 74, 78, 79, 88, 91, 105, 106, 110, 124 |
| 5 interacts with | 4, 6, 10, 14, 17, 21, 24, 46, 102, 113, 116, 118, 131, 133, 140 |
| 6 interacts with | 2, 4, 5, 20, 21, 53, 58, 61, 63, 82, 102, 111, 117, 130, 134, 135 |
| 7 interacts with | 20, 31, 34, 53, 57, 58, 59, 80, 85, 86, 94, 105, 106, 123, 124, 125 |
| 8 interacts with | 1, 9, 14, 15, 21, 22, 25, 27, 48, 58, 59, 61, 62, 63, 64, 65, 89, 95, 96, 99, 111, 112, 114, 115, 116, 121, 129, 136, 140, 141 |
| 9 interacts with | 1, 8, 11, 12, 13, 15, 17, 18, 20, 21, 28, 29, 36, 43, 49, 56, 62, 64, 80, 81, 101, 113, 118, 124, 129, 136, 140, 141 |
| 10 interacts with | 5, 13, 14, 15, 18, 24, 26, 65, 68, 93, 102, 134 |
| 11 interacts with | 9, 12, 64, 95, 96, 114, 133, 134 |
| 12 interacts with | 1, 3, 9, 11, 17, 18, 19, 25, 26, 28, 34, 36, 41, 43, 49, 56, 62, 63, 76, 80, 81, 85, 86, 87, 90, 91, 93, 121, 122, 129, 140, 141 |
| 13 interacts with | 1, 3, 9, 10, 17, 20, 25, 28, 33, 34, 36, 37, 41, 45, 56, 62, 68, 79, 80, 81, 83, 86, 91, 94, 101, 106, 108, 121, 122, 129, 137, 140, 141 |
| 14 interacts with | 1, 5, 8, 10, 15, 19, 20, 21, 28, 30, 40, 43, 44, 47, 54, 62, 63, 64, 65, 86, 97, 121, 129, 130, 133, 138, 141 |
| 15 interacts with | 8, 9, 10, 14, 18, 21, 22, 37, 39, 41, 44, 45, 46, 58, 59, 61, 62, 63, 64, 65, 66, 95, 96, 97, 98, 112, 116, 125, 127, 128, 129, 130, 132, 133, 135, 137, 138, 141 |
| 16 interacts with | 27, 29, 34, 68, 78, 79, 82, 88, 95, 101, 114, 117, 119, 122 |
| 17 interacts with | 3, 4, 5, 9, 12, 13, 20, 23, 27, 37, 38, 43, 49, 65, 69, 80, 81, 86, 89, 101, 110, 115, 116, 117, 118, 126, 129, 135 |
| 18 interacts with | 9, 10, 12, 15, 19, 26, 28, 31, 33, 42, 43, 44, 47, 48, 49, 60, 65, 69, 70, 74, 77, 79, 85, 97, 98, 103, 110, 140, 141 |
| 19 interacts with | 12, 14, 18, 22, 26, 28, 32, 33, 36, 37, 38, 41, 45, 49, 69, 71, 86, 104, 106, 107, 110, 118, 126, 140 |
| 20 interacts with | 6, 9, 13, 14, 17, 24, 29, 30, 36, 37, 43, 54, 64, 68, 80, 84, 89, 102, 116, 117, 129, 131, 133, 140 |
| 21 interacts with | 1, 5, 6, 8, 9, 14, 15, 24, 61, 63, 89, 95, 96, 111, 112, 113, 115, 116, 137, 139, 140, 141 |
| 22 interacts with | 8, 15, 19, 31, 32, 33, 36, 42, 44, 47, 49, 60, 61, 64, 69, 71, 74, 97, 98, 104, 107, 110, 127, 140 |
| 23 interacts with | 4, 17, 31, 34, 62, 63, 71, 76, 78, 79, 82, 83, 93, 95, 100, 101, 105, 115, 116, 119, 126, 132, 137 |
| 24 interacts with | 5, 10, 20, 21, 38, 82, 93, 100, 101, 102, 108, 115, 130, 133, 135, 140, 141 |
| 25 interacts with | 8, 12, 13, 26, 27, 36, 62, 81, 90, 92, 111, 114, 116, 120 |
| 26 interacts with | 2, 3, 10, 12, 18, 19, 25, 29, 31, 33, 34, 41, 53, 56, 58, 62, 67, 68, 76, 85, 90, 91, 92, 93, 108, 113, 122, 123, 124, 130 |
| 27 interacts with | 8, 16, 17, 25, 29, 62, 68, 81, 86, 88, 90, 92, 113, 114, 122, 130 |
| 28 interacts with | 1, 9, 12, 13, 14, 18, 19, 29, 31, 33, 34, 35, 36, 37, 38, 42, 45, 49, 50, 54, 55, 56, 62, 74, 92, 103, 106, 107, 108, 109, 110, 118, 127, 129, 131 |
| 29 interacts with | 1, 2, 4, 9, 16, 20, 26, 27, 28, 41, 67, 71, 81, 85, 88, 92, 101, 119, 122, 124 |
| 30 interacts with | 7, 14, 20, 31, 33, 35, 40, 47, 63, 95, 97, 98, 107, 126, 127, 129, 130, 131, 132, 133, 139 |
| 31 interacts with | 7, 18, 22, 23, 26, 28, 30, 33, 34, 35, 37, 40, 43, 44, 49, 50, 52, 54, 59, 60, 80, 89, 94, 98, 106, 107, 109, 128, 131, 132 |
| 32 interacts with | 2, 4, 19, 22, 34, 42, 43, 46, 48, 52, 54, 60, 61, 63, 65, 69, 70, 71, 73, 74, 75, 104, 105, 107, 109, 110, 122, 129 |
| 33 interacts with | 13, 18, 19, 22, 26, 28, 30, 31, 34, 35, 36, 41, 54, 56, 59, 74, 78, 80, 90, 91, 92, 94, 105, 107, 118, 122, 123, 124, 136 |
| 34 interacts with | 7, 12, 13, 16, 23, 26, 28, 31, 32, 33, 41, 54, 56, 59, 74, 78, 80, 90, 91, 92, 94, 105, 107, 118, 122, 123, 124, 136 |
| 35 interacts with | 28, 30, 31, 33, 39, 42, 43, 46, 61, 79, 104, 118, 137 |
| 36 interacts with | 9, 12, 13, 19, 20, 22, 25, 28, 33, 38, 40, 41, 43, 45, 52, 54, 61, 68, 80, 81, 86, 94, 106, 110, 136 |
| 37 interacts with | 13, 15, 17, 19, 20, 28, 31, 38, 43, 44, 49, 50, 72, 76, 97, 103, 128, 133, 140 |
| 38 interacts with | 17, 19, 24, 28, 36, 37, 40, 42, 43, 44, 50, 52, 58, 61, 68, 76, 78, 79, 94, 97, 106, 128 |
| 39 interacts with | 15, 33, 35, 44, 48, 62, 69, 70, 72, 75, 97, 104, 118, 127, 134, 137, 138 |
| 40 interacts with | 14, 30, 31, 33, 36, 38, 42, 44, 48, 69, 70, 97, 104, 107, 118, 125, 127, 134, 137, 138 |
| 41 interacts with | 12, 13, 15, 19, 26, 29, 33, 34, 36, 44, 48, 51, 65, 69, 70, 71, 72, 92, 98, 104, 107, 118, 122, 125, 127, 138 |
| 42 interacts with | 18, 22, 28, 32, 33, 35, 38, 40, 43, 48, 49, 50, 57, 69, 104, 105, 107, 110, 118, 137 |
| 43 interacts with | 9, 12, 14, 17, 18, 20, 31, 32, 33, 35, 36, 37, 38, 42, 48, 51, 60, 64, 69, 71, 86, 101, 104, 107, 109, 119, 129, 140 |
| 44 interacts with | 14, 15, 18, 22, 31, 37, 38, 39, 40, 41, 51, 52, 60, 62, 87, 97, 98, 110 |
| 45 interacts with | 4, 13, 15, 19, 28, 36, 48, 54, 65, 69, 70, 71, 73, 74, 78, 79, 91, 104, 105, 106, 110, 118, 125, 127, 130, 138 |
| 46 interacts with | 5, 15, 32, 33, 35, 47, 66, 106, 107, 118, 130 |
| 47 interacts with | 14, 18, 22, 30, 33, 46, 62, 107, 118, 130 |
| 48 interacts with | 1, 8, 18, 32, 33, 39, 40, 41, 42, 43, 45, 52, 63, 71, 75, 85, 86, 97, 99, 105, 107, 109, 110, 119, 129, 130, 141 |

**Notes on "Notes on the Synthesis of Form"**

| | |
|---|---|
| 49 interacts with | 9, 12, 17, 18, 19, 22, 28, 31, 37, 42, 51, 64, 68, 86, 97, 107, 110, 117, 118, 128, 129, 130, 132, 133, 138, 140 |
| 50 interacts with | 28, 31, 37, 38, 42, 52, 54, 60, 76, 77, 85, 87, 94, 103 |
| 51 interacts with | 33, 41, 43, 44, 49, 53, 54, 59, 69, 77, 104, 107, 118, 127, 136 |
| 52 interacts with | 2, 31, 32, 36, 38, 44, 48, 50, 53, 54, 59, 71, 91, 104, 106, 107, 136 |
| 53 interacts with | 6, 7, 26, 51, 52, 56, 57, 59, 60, 66, 72, 118, 126, 127, 137 |
| 54 interacts with | 14, 20, 28, 31, 32, 33, 34, 36, 45, 50, 51, 52, 56, 57, 59, 71, 80, 91, 94, 106, 107, 110, 115 |
| 55 interacts with | 28, 67, 68, 71, 80, 119, 123, 124 |
| 56 interacts with | 4, 9, 12, 13, 26, 28, 34, 53, 54, 57, 59, 76, 78, 80, 85, 86, 92, 102, 123, 124 |
| 57 interacts with | 7, 42, 53, 54, 56, 59, 60, 70, 86, 94, 117, 118, 123, 126, 127, 137 |
| 58 interacts with | 6, 7, 8, 15, 26, 38, 65, 72, 76, 78, 93, 96, 98, 99, 125, 127, 130, 138 |
| 59 interacts with | 7, 8, 15, 31, 34, 51, 52, 53, 54, 57, 58, 60, 65, 66, 72, 96, 98, 99, 125, 127, 130, 138 |
| 60 interacts with | 18, 22, 31, 32, 43, 44, 50, 53, 57, 59, 91, 94, 97, 98, 103, 131 |
| 61 interacts with | 1, 6, 8, 15, 21, 22, 32, 35, 36, 38, 63, 74, 86, 95, 96, 97, 98, 99, 105, 108, 109, 110, 119, 120, 127, 131, 139, 140, 141 |
| 62 interacts with | 8, 9, 12, 13, 14, 15, 23, 25, 26, 27, 28, 39, 44, 47, 65, 66, 72, 85, 86, 87, 89, 93, 114, 115, 116, 119, 127, 130, 132, 138, 141 |
| 63 interacts with | 4, 6, 8, 12, 14, 15, 21, 23, 30, 32, 48, 61, 64, 65, 66, 68, 70, 71, 72, 75, 86, 93, 96, 99, 100, 116, 119, 127, 129, 130, 132, 133, 134, 136, 138, 140, 141 |
| 64 interacts with | 8, 9, 11, 14, 15, 20, 22, 43, 49, 63, 81, 85, 86, 95, 99, 100, 101, 109, 112, 113, 127, 130, 133, 136, 139 |
| 65 interacts with | 8, 10, 14, 15, 17, 18, 32, 41, 45, 58, 59, 62, 63, 66, 72, 84, 99, 111, 114, 116, 127, 130, 133, 134, 138, 139, 141 |
| 66 interacts with | 15, 46, 53, 59, 62, 63, 65, 68, 70, 71, 75, 93, 130, 132, 133, 137, 139, 141 |
| 67 interacts with | 1, 26, 29, 55, 76, 86, 92, 122, 123 |
| 68 interacts with | 1, 10, 13, 16, 20, 26, 27, 36, 38, 49, 55, 63, 66, 71, 86, 94, 101, 109, 110, 114, 119, 124, 129, 131, 132, 141 |
| 69 interacts with | 17, 18, 19, 22, 32, 33, 39, 40, 41, 42, 43, 45, 51, 74, 75, 92, 104, 105, 107, 132 |
| 70 interacts with | 1, 18, 32, 33, 39, 40, 41, 45, 57, 63, 66, 71, 72, 73, 86, 104, 110, 131, 132 |
| 71 interacts with | 2, 4, 19, 22, 23, 29, 32, 33, 41, 43, 45, 48, 52, 54, 55, 63, 66, 68, 70, 73, 75, 76, 79, 88, 98, 104, 105, 107, 108, 109, 110, 120, 129, 131, 132, 133 |
| 72 interacts with | 33, 37, 39, 41, 53, 58, 59, 62, 63, 65, 70, 104, 128, 130, 131 |
| 73 interacts with | 32, 45, 70, 71, 78, 91, 104, 105, 108, 109, 110 |
| 74 interacts with | 4, 18, 22, 28, 32, 33, 34, 45, 61, 69, 105, 107, 109, 110, 127 |
| 75 interacts with | 32, 33, 39, 48, 63, 66, 69, 71, 98, 100, 104, 107, 123, 124, 133 |
| 76 interacts with | 3, 12, 23, 26, 37, 38, 50, 56, 58, 67, 71, 85, 87, 90, 91, 92, 95, 98, 101, 108, 113, 120, 122, 123, 124, 127 |
| 77 interacts with | 1, 18, 50, 51, 79, 83, 86, 90, 93, 103 |
| 78 interacts with | 3, 4, 16, 23, 34, 38, 45, 56, 58, 73, 79, 85, 86, 101, 105, 130 |
| 79 interacts with | 3, 4, 13, 16, 18, 23, 35, 38, 45, 71, 77, 78, 86, 88, 90, 104, 105, 111, 116, 124, 127, 130 |
| 80 interacts with | 7, 9, 12, 13, 17, 20, 31, 34, 36, 54, 55, 56, 86, 94, 103, 106, 123, 136 |
| 81 interacts with | 9, 12, 13, 17, 25, 27, 29, 36, 64, 82, 83, 85, 86, 92, 93, 113, 114, 119, 122, 133, 136 |
| 82 interacts with | 6, 16, 23, 24, 81, 111, 113, 115 |
| 83 interacts with | 13, 23, 77, 81, 85, 86, 101 |
| 84 interacts with | 20, 65, 120, 127, 131, 132, 134, 135 |
| 85 interacts with | 7, 12, 18, 26, 29, 48, 50, 56, 62, 64, 76, 78, 81, 83, 86, 87, 93, 108, 136 |
| 86 interacts with | 1, 3, 7, 12, 13, 14, 17, 19, 27, 36, 43, 48, 49, 56, 57, 61, 62, 63, 64, 67, 68, 70, 77, 78, 79, 80, 81, 83, 85, 103, 111, 117, 119, 120, 121, 123, 124, 125, 140, 141 |
| 87 interacts with | 12, 44, 50, 62, 76, 85, 90, 91, 93, 95, 100, 128 |
| 88 interacts with | 4, 16, 27, 29, 71, 79, 114, 123 |
| 89 interacts with | 8, 17, 20, 21, 31, 62, 100, 130, 138, 141 |
| 90 interacts with | 12, 25, 26, 27, 33, 34, 76, 77, 79, 87, 91, 93, 113, 114, 121, 124, 132 |
| 91 interacts with | 4, 12, 13, 26, 33, 34, 45, 52, 54, 60, 73, 76, 87, 90, 103, 105, 121, 132 |
| 92 interacts with | 25, 26, 27, 28, 29, 34, 41, 56, 67, 69, 76, 81, 114, 122, 123, 124, 132 |
| 93 interacts with | 10, 12, 23, 24, 26, 62, 63, 66, 77, 81, 87, 90, 116, 130, 132, 137, 141 |
| 94 interacts with | 13, 31, 34, 36, 38, 50, 54, 55, 57, 60, 68, 80, 103, 106, 119, 136 |
| 95 interacts with | 8, 11, 15, 16, 21, 23, 30, 61, 64, 76, 87, 102, 112, 117, 119, 121, 130, 132, 133, 135, 139, 141 |
| 96 interacts with | 8, 11, 15, 21, 58, 59, 61, 63, 97, 102, 119, 121, 130, 132, 133, 139, 141 |
| 97 interacts with | 14, 15, 18, 22, 30, 37, 38, 39, 40, 44, 48, 49, 60, 61, 96, 98, 119, 132, 133, 135 |
| 98 interacts with | 2, 15, 18, 22, 30, 31, 41, 44, 58, 59, 60, 61, 71, 75, 76, 97, 109, 110, 119, 120, 121, 132, 133, 139 |
| 99 interacts with | 8, 48, 58, 59, 61, 63, 64, 65, 131, 132, 133, 138 |
| 100 interacts with | 23, 24, 63, 64, 75, 87, 89, 101, 112, 113, 115, 121, 126, 130, 132, 133, 135, 141 |
| 101 interacts with | 1, 3, 9, 13, 16, 17, 23, 24, 29, 43, 64, 68, 76, 78, 83, 100, 102, 112, 113, 117, 119, 122, 133 |
| 102 interacts with | 2, 5, 6, 10, 20, 24, 56, 95, 96, 101, 115 |
| 103 interacts with | 3, 18, 28, 37, 50, 60, 77, 80, 86, 91, 94 |
| 104 interacts with | 19, 22, 32, 33, 35, 39, 40, 41, 42, 43, 45, 51, 52, 69, 70, 71, 72, 73, 75, 79, 105, 107, 109 |
| 105 interacts with | 2, 4, 7, 23, 32, 33, 34, 42, 45, 48, 61, 69, 71, 73, 74, 78, 79, 91, 104, 106, 110, 119, 137 |
| 106 interacts with | 1, 4, 7, 13, 19, 28, 31, 36, 38, 45, 46, 52, 54, 80, 94, 105, 129, 136 |
| 107 interacts with | 19, 22, 28, 30, 31, 32, 33, 34, 40, 41, 42, 43, 46, 47, 48, 49, 51, 52, 54, 69, 71, 74, 75, 104, 110, 122, 136 |
| 108 interacts with | 13, 24, 26, 28, 61, 73, 76, 85, 109, 110 |
| 109 interacts with | 28, 31, 32, 43, 48, 61, 64, 68, 71, 73, 74, 98, 104, 108, 110 |

**Notes on "Notes on the Synthesis of Form"**

| | |
|---|---|
| 110 interacts with | 4, 17, 18, 19, 22, 28, 32, 33, 36, 42, 43, 44, 45, 48, 49, 54, 61, 68, 70, 71, 73, 74, 98, 105, 107, 108, 109, 137 |
| 111 interacts with | 6, 8, 21, 25, 65, 79, 82, 86, 113, 115, 116, 117, 120, 130, 132, 134 |
| 112 interacts with | 8, 15, 21, 64, 95, 100, 101, 130, 133, 139, 141 |
| 113 interacts with | 1, 5, 9, 21, 26, 27, 64, 76, 81, 82, 90, 100, 101, 111, 114, 117, 119, 124 |
| 114 interacts with | 8, 11, 16, 25, 27, 62, 65, 68, 81, 88, 90, 92, 113, 117, 123, 127, 130, 132 |
| 115 interacts with | 8, 17, 21, 23, 24, 54, 62, 82, 100, 102, 111, 127, 132, 137, 140, 141 |
| 116 interacts with | 5, 8, 15, 17, 20, 21, 23, 25, 62, 63, 65, 79, 111, 117, 121, 127, 128, 131, 132, 135, 137 |
| 117 interacts with | 6, 16, 17, 20, 49, 57, 86, 95, 101, 111, 113, 114, 116, 121, 123, 124, 125, 133, 135, 137 |
| 118 interacts with | 5, 9, 17, 19, 28, 33, 34, 35, 39, 40, 41, 42, 45, 46, 47, 49, 51, 53, 57, 126, 127, 130, 131, 134 |
| 119 interacts with | 3, 16, 23, 29, 48, 55, 61, 62, 63, 68, 81, 86, 94, 95, 96, 97, 98, 101, 105, 113, 136 |
| 120 interacts with | 25, 61, 71, 76, 84, 86, 98, 111, 121, 126, 132, 133, 139 |
| 121 interacts with | 8, 12, 13, 14, 86, 90, 91, 95, 96, 98, 100, 116, 117, 120, 123, 124, 125, 127, 132, 133, 139 |
| 122 interacts with | 12, 13, 16, 26, 27, 29, 32, 33, 34, 41, 67, 76, 92, 101, 107, 123 |
| 123 interacts with | 2, 7, 26, 34, 55, 56, 57, 67, 75, 76, 80, 86, 88, 92, 114, 117, 121, 122, 127, 137 |
| 124 interacts with | 1, 4, 7, 9, 26, 29, 34, 55, 56, 68, 75, 76, 79, 86, 90, 92, 113, 117, 121, 137 |
| 125 interacts with | 7, 15, 40, 41, 45, 58, 59, 86, 117, 121 |
| 126 interacts with | 17, 19, 30, 33, 53, 57, 100, 118, 120, 133 |
| 127 interacts with | 15, 22, 28, 30, 33, 39, 40, 41, 45, 51, 53, 57, 58, 59, 61, 62, 63, 64, 65, 74, 76, 79, 84, 114, 115, 116, 118, 121, 123, 132, 135 |
| 128 interacts with | 15, 31, 33, 37, 38, 49, 72, 87, 116, 138, 140 |
| 129 interacts with | 8, 9, 12, 13, 14, 15, 17, 20, 28, 30, 32, 43, 48, 49, 63, 68, 71, 106, 131, 140 |
| 130 interacts with | 6, 10, 14, 15, 24, 26, 27, 30, 45, 46, 47, 48, 49, 58, 59, 62, 63, 64, 65, 66, 72, 78, 79, 89, 93, 95, 96, 100, 111, 112, 114, 118, 134, 137, 141 |
| 131 interacts with | 5, 20, 28, 30, 31, 60, 61, 68, 70, 71, 72, 84, 99, 116, 118, 129, 135 |
| 132 interacts with | 15, 23, 30, 31, 49, 62, 63, 66, 68, 69, 70, 71, 84, 90, 91, 92, 93, 95, 96, 97, 98, 99, 100, 111, 114, 115, 116, 120, 121, 127 |
| 133 interacts with | 2, 5, 10, 11, 14, 15, 20, 24, 30, 37, 49, 63, 64, 65, 66, 71, 75, 81, 95, 96, 97, 98, 99, 100, 101, 112, 117, 120, 121, 126, 134, 136, 139, 140 |
| 134 interacts with | 6, 10, 11, 33, 39, 40, 63, 65, 84, 111, 118, 130, 133 |
| 135 interacts with | 6, 15, 17, 24, 84, 95, 97, 100, 116, 117, 127, 131, 137 |
| 136 interacts with | 8, 9, 34, 36, 51, 52, 63, 64, 80, 81, 85, 94, 106, 107, 119, 133, 140 |
| 137 interacts with | 13, 15, 21, 23, 33, 35, 39, 40, 42, 53, 57, 66, 93, 105, 110, 115, 116, 117, 123, 124, 130, 135, 140 |
| 138 interacts with | 14, 15, 33, 39, 40, 41, 45, 49, 58, 59, 62, 63, 65, 89, 128, 140, 141 |
| 139 interacts with | 21, 30, 61, 64, 65, 66, 95, 96, 98, 112, 120, 121, 133 |
| 140 interacts with | 1, 5, 8, 9, 12, 13, 18, 19, 20, 21, 22, 24, 37, 43, 49, 61, 63, 86, 115, 128, 129, 133, 136, 137, 138, 141 |
| 141 interacts with | 1, 8, 9, 12, 13, 14, 15, 18, 21, 24, 48, 61, 62, 63, 65, 66, 68, 86, 89, 93, 95, 96, 100, 112, 115, 130, 138, 140 |

There are 50 errors in this table, each of the form of an asymmetry: for some $i$ and $j$, "$i$ interacts with $j$" is included but "$j$ interacts with $i$" is not also included. Here are all the 1-way interactions; the arrows indicate the specified direction of interaction:

| | |
|---|---|
| 3 interacts with: | ←86, →88 |
| 4 interacts with: | ←23 |
| 7 interacts with: | →20, ←30, →94 |
| 10 interacts with: | ←130, ←133 |
| 23 interacts with: | →126 |
| 33 interacts with: | ←39, ←40, ←42, ←43, ←46, ←47, ←48, ←51, →56, →59, ←69, ←70, ←71, ←72, ←75, →78, →80, →92, →94, ←104, ←110, →123, →124, ←126, ←127, ←128, ←134, →136, ←137, ←138 |
| 43 interacts with: | ←110, →119 |
| 55 interacts with: | ←94 |
| 56 interacts with: | →59 |
| 58 interacts with: | ←59, →93 |
| 71 interacts with: | →108 |
| 81 interacts with: | →122 |
| 85 interacts with: | →93 |
| 93 interacts with: | →116 |
| 99 interacts with: | →138 |

Notice that of the 50 errors in the interactions table, 30 involve requirement 33: "Fertile land to be used to best advantage."

The HIDECS 2 report states that errors like this are handled by a consistency checker on input from punched cards, as discussed in Appendix F.

## C  Alexander's Decomposition in "Notes"

The following are the elements in the Indian Village decomposition in "Notes."

| | |
|---|---|
| A contains requirements | 7, 31, 34, 36, 37, 38, 50, 52, 53, 54, 55, 57, 59, 60, 72, 77, 80, 91, 94, 103, 106, 125, 126, 128, 136 |
| B contains requirements | 18, 19, 22, 28, 30, 32, 33, 35, 39, 40, 41, 42, 43, 44, 45, 46, 47, 48, 49, 51, 61, 69, 70, 71, 73, 74, 75, 97, 98, 104, 105, 107, 108, 109, 110, 118, 127, 131, 138 |

**Notes on "Notes on the Synthesis of Form"**

| | |
|---|---|
| C contains requirements | 5, 6, 8, 10, 11, 14, 15, 20, 21, 24, 58, 63, 64, 65, 66, 84, 89, 93, 95, 96, 99, 100, 102, 111, 112, 115, 116, 117, 120, 121, 129, 130, 132, 133, 134, 135, 137, 139, 140, 141 |
| D contains requirements | 1, 2, 3, 4, 9, 12, 13, 16, 17, 23, 25, 26, 27, 29, 56, 62, 67, 68, 76, 78, 79, 81, 82, 83, 85, 86, 87, 88, 90, 92, 101, 113, 114, 119, 122, 123, 124 |
| A1 contains requirements | 7, 53, 57, 59, 60, 72, 125, 126, 128 |
| A2 contains requirements | 31, 34, 36, 52, 54, 80, 94, 106, 136 |
| A3 contains requirements | 37, 38, 50, 55, 77, 91, 103 |
| B1 contains requirements | 39, 40, 41, 44, 51, 118, 127, 131, 138 |
| B2 contains requirements | 30, 35, 46, 47, 61, 97, 98 |
| B3 contains requirements | 18, 19, 22, 28, 33, 42, 43, 49, 69, 74, 107, 110 |
| B4 contains requirements | 32, 45, 48, 70, 71, 73, 75, 104, 105, 108, 109 |
| C1 contains requirements | 8, 10, 11, 14, 15, 58, 63, 64, 65, 66, 93, 95, 96, 99, 100, 112, 121, 130, 132, 133, 134, 139, 141 |
| C2 contains requirements | 5, 6, 20, 21, 24, 84, 89, 102, 111, 115, 116, 117, 120, 129, 135, 137, 140 |
| D1 contains requirements | 26, 29, 56, 67, 76, 85, 87, 90, 92, 122, 123, 124 |
| D2 contains requirements | 1, 9, 12, 13, 25, 27, 62, 68, 81, 86, 113, 114 |
| D3 contains requirements | 2, 3, 4, 16, 17, 23, 78, 79, 82, 83, 88, 101, 119 |

## D  The Goodness Measures

For these goodness functions I use Alexander's names for variables and inputs where possible.

### D.1  HIDECS2-Actual

Figure 7 shows the goodness function used by HIDECS 2 as described in the HIDECS 2 Report [16]. It is not clear whether this is the function actually used in the program that generated the decomposition in "Notes," because in another paper, "The Determination of Components for an Indian Village" [4], Alexander shows a different goodness measure (HIDECS2-Decomp, below) and states that it was used to create the decomposition listed in that paper, and that decomposition is identical to the decomposition in "Notes."

The actual quantity to be minimized is STR, but back in 1960, doing a square root was not so fast on a computer; so instead, the following order-preserving variant is used:

> Given $nbit, total, M, N$:
>
> $$nbit = \text{number of nodes} \tag{3}$$
> $$total = \text{number of interaction links} \tag{4}$$
> $$M, N = \text{partition of the nodes} \tag{5}$$
> $$m = |M| \tag{6}$$
> $$n = |N| \tag{7}$$
> $$RR = \sum_{i \in M, j \in N} v_{ij} \tag{8}$$
> $$nsq1 = \frac{1}{2}nbit(nbit-1) = \text{maximum number of possible links} \tag{9}$$
> $$\text{STR} = \frac{RR - \left(\frac{total}{nsq1}\right)mn}{\sqrt{mn(nsq1-mn)}} \tag{10}$$

**Figure 7** HIDECS2-Actual

$$nom = RR - \left(\frac{total}{nsq1}\right)mn \tag{11}$$
$$denom = mn(nsq1-mn) \tag{12}$$
$$\text{INFO} = signum(nom)\frac{nom^2}{denom} \tag{13}$$

and the result is INFO. Basically, the fraction is squared but the sign is preserved. As Alexander wrote it in the HIDECS 2 report, INFO = (STR)(|STR|).[21]

**The Intuition**: HIDECS2-Actual computes the goodness of a bifurcation of a set of nodes with respect to a given set of interaction links; this measure is to be minimized. Essentially it computes the coupling between sets in a partition. The numerator measures how much the actual number of links cut by the partition differs from the theoretical expected number cut. The product, $mn$, is the number of possible links between $M$ and $N$, and

$$\frac{total}{nsq1}$$

is the ratio of specified interaction links to the maximum number of possible links. Thus

$$\left(\frac{total}{nsq1}\right)mn$$

---

[21] On page 25 of the scanned photocopy sent to me (page 25 is in Appendix N), this equation has a large, penciled-in ? next to it. I took the formulation as evidence of some hackerish sophistication, either by Alexander or Manheim.

is the expected number of links cut by the partition. When the numerator is negative, there are fewer links cut than what is expected; if positive, there are more. The smallest the numerator can achieve occurs when $RR = 0$.

The denominator corrects for a bias Alexander wants to avoid—a partition into a small set and a large one, that is, an imbalance where $m$ is small and $n$ is large. For a designer such a partition would not be very helpful. In the denominator, $nsq1 - mn$ is the maximum number of possible links that are contained within either $M$ or $N$. When $mn$ is small, $nsq1 - mn$ is large, and as $m$ goes from 1 up to its maximum, $mn$ looks like a frown and $nsq1 - mn$ like a smile; the overall effect of the denominator is to rebalance the goodness measure: the shape of $mn(nsq1 - mn)$ as the size of $M$ rises from 1 to $nbit - 1$ is like an upside down parabola with a dimple at the top:

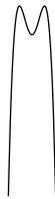

This has the effect of favoring partitions that are about the same size. Alexander put it this way in HIDECS 2 (page E2):

> *As the measure* [roughly, *RR*] *stands, however, it is biased toward strongly assymmetrical* [sic][22] *partitions, in which the product MN is small—e.g., where M is small and N large. We normalize the measure by subtracting the expected value of RR and dividing by the square root of its variance. The normalized redundancy is . . . .*

The fact that Alexander noticed that some goodness measures would tend to favor unbalanced partitions indicates to me that he had tried several intuitive measures—I had the same experience.

(Confession: I could not figure out how "dividing by the square root of its variance" turned into the denominator shown. "Variance" of what? My perplexity is described in Appendix O.)

### D.2 HIDECS2-Decomp

Figure 8 shows the goodness measure shown in "The Determination of Components for an Indian Village" [4]. In that paper Alexander states the following:

> *minimization according to this function has been programmed for the IBM 7090.*[23]
> *It is this function which gave the decomposition of the village problem that follows.*

The decomposition that followed was exactly the one in "Notes."

---

[22] Spelling.
[23] This IBM 7090 was located at the Smithsonian Astrophysical Observatory in Cambridge, Massachusetts.

$$m = \text{total number of variables} \tag{14}$$
$$a = \text{number of variables in one subsystem} \tag{15}$$
$$b = \text{number of variables in the other subsystem} \tag{16}$$
$$l = \text{total number of links} \tag{17}$$
$$l_a = \text{number of links entirely within the first subsystem} \tag{18}$$
$$l_b = \text{number of links entirely within the other subsystem} \tag{19}$$
$$\text{SCORE} = \frac{(l - l_a - l_b)(\frac{1}{2}m(m-1)) - lab}{\sqrt{ab(\frac{1}{2}m(m-1) - ab)}} \tag{20}$$

**Figure 8** HIDECS2-Decomp

$$m = nbit \tag{21}$$
$$a = m \tag{22}$$
$$b = n \tag{23}$$
$$ab = mn \tag{24}$$
$$l = total \tag{25}$$
$$l - l_a - l_b = RR \tag{26}$$
$$\frac{1}{2}m(m-1) = nsq1 \tag{27}$$
$$\text{SCORE} = \frac{RR \cdot nsq1 - l \cdot mn}{\sqrt{mn(nsq1 - mn)}} \tag{28}$$
$$= nsq1 \cdot \text{STR} \tag{29}$$

**Figure 9** HIDECS2-Decomp vs HIDECS2-Actual

Some of this should look familiar. Looking at HIDECS2-Actual, we can observe the correspondence shown in Figure 9.

On page 190 of "Notes," Alexander talks about choosing a constant to simplify things: "the...redundancy of a partition $\pi$, is"

$$R(\pi) = \frac{\text{constant}(l^\pi - \frac{ll_0^\pi}{l_0})}{\sqrt{\frac{ll_0^\pi(l_0 - l_0^\pi)}{l_0(l_0 - 1)}}}$$

"...choose the constant to make this" equal to

$$\frac{l_0 l^\pi - ll_0^\pi}{\sqrt{l_0^\pi(l_0 - l_0^\pi)}}$$

where

**Notes on "Notes on the Synthesis of Form"**

$$R(\pi) = \frac{\frac{1}{2}m(m-1)\sum_{\pi} v_{ij} - l\sum_{\pi} S_\alpha S_\beta}{\left[(\sum_{\pi} S_\alpha S_\beta)(\frac{1}{2}m(m-1) - \sum_{\pi} S_\alpha S_\beta)\right]^{\frac{1}{2}}}$$

where

$$m = \text{number of nodes} \tag{33}$$
$$l = \text{total number of links} \tag{34}$$
$$\sum_{\pi} v_{ij} = \text{number of links cut by the partition} \tag{35}$$
$$S_x = \text{the number of nodes in partition } x \tag{36}$$
$$\frac{1}{2}m(m-1) = \text{the maximum number of links possible} \tag{37}$$

■ **Figure 10** HIDECS2-Notes

$$l_0 = \frac{1}{2}m(m-1) \tag{30}$$
$$l^\pi = \sum_{\pi} v_{ij} \tag{31}$$
$$l_0^\pi = \sum_{\pi} S_\alpha S_\beta \tag{32}$$

Note that even though the different variants don't compute the same values, ordering is preserved.

### D.3 HIDECS2-Notes

The goodness measure described in "Notes" is shown in Figure 10. It is the straightforward generalization of HIDECS2-Decomp to more than two sets.

For partitions of a set into two parts, HIDECS2-Decomp and HIDECS2-Notes compute the same value.

### D.4 HIDECS2-rpg

To explore the space of decompositions, I came up with a goodness measure that seemed aimed directly at cohesion / coupling, as shown in Figure 11.

This might look complicated, but its narrative is simple.

**The Intuition**: The variables $f_1$ and $f_3$ measure cohesion, and $f_2$ coupling. We minimize RSCORE. The variable $l_a$ is the number of links entirely within the partition referred to as $a$, and $m_a$ is the theoretical maximum for such links; and so $f_1$ is 0 when the number of links that lie completely within the first partition is the maximum

$$m = \text{total number of variables} \tag{38}$$
$$a = \text{number of variables in one subsystem} \tag{39}$$
$$b = \text{number of variables in the other subsystem} \tag{40}$$
$$l = \frac{1}{2}m(m-1) \tag{41}$$
$$l_a = \text{number of links entirely within the first subsystem} \tag{42}$$
$$l_b = \text{number of links entirely within the other subsystem} \tag{43}$$
$$m_a = \frac{a(a-1)}{2} \tag{44}$$
$$m_b = \frac{b(b-1)}{2} \tag{45}$$
$$l_{ab} = l - l_a - l_b \tag{46}$$
$$m_{ab} = m_a m_b \tag{47}$$
$$f_1 = \begin{cases} 0 & \text{if } m_a = 0 \\ 1 - \frac{l_a}{m_a} & \text{otherwise} \end{cases} \tag{48}$$
$$f_2 = \begin{cases} 0 & \text{if } m_{ab} = 0 \\ \frac{l_{ab}}{m_{ab}} & \text{otherwise} \end{cases} \tag{49}$$
$$f_3 = \begin{cases} 0 & \text{if } m_b = 0 \\ 1 - \frac{l_b}{m_b} & \text{otherwise} \end{cases} \tag{50}$$
$$f_4 = 1 - \left(\frac{a-b}{a+b}\right)^2 \tag{51}$$
$$\text{RSCORE} = \frac{f_1 + f_2 + f_3}{f_4} \tag{52}$$

■ **Figure 11** HIDECS2-rpg

possible ($\frac{l_a}{m_a} = 1$), and 1 when none of them are. Similarly for $f_3$. Variable $f_2$ is 0 when there are no links between the two partitions, and 1 if all of them are—because $l_{ab}$ is the number of links between the partitions, and $m_{ab}$ is the theoretical maximum.

Therefore, $f_1 + f_2 + f_3$ is minimal when there is good cohesion and weak coupling. But there is that problem of tending to favor lopsided partitions, as Alexander mentions. The variable $f_4$ then measures the balance: it is maximum when the partitions are the same size and minimum with one of them has only 1 element. It is an upside-down curve, and though it is not as sharp as the one in HIDECS2-Actual, it has a similar shape:

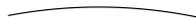

**Notes on "Notes on the Synthesis of Form"**

### D.5 HIDECS3-BLDUP

The goodness measure used for BLDUP, which is described in the HIDECS 3 report, is derived from HIDECS2-Actual (Equation 10):

$$\text{STR} = \frac{RR - \left(\frac{total}{nsq1}\right)mn}{\sqrt{mn(nsq1 - mn)}}$$

If we replace *RR* by

$$\sum_\pi v_{i,j}$$

and *mn* by

$$\sum_\pi S_\alpha S_\beta$$

we get

$$\text{STR}_2 = \frac{\sum_\pi v_{ij} - \left(\frac{total}{nsq1}\right)\sum_\pi S_\alpha S_\beta}{\sqrt{(\sum_\pi S_\alpha S_\beta)(nsq1 - \sum_\pi S_\alpha S_\beta)}} = \frac{nom}{denom}$$

and then the goodness measure used for BLDUP is

$$\text{INFO}_2 = signum(nom)\frac{nom^2}{denom}$$

Note that if we take

$$l = total = \text{total number of links} \tag{53}$$

$$nsq1 = \frac{1}{2}m(m-1) \tag{54}$$

then HIDECS2-Notes is $nsq1 \cdot \text{STR}_2$.

### D.6 HIDECS3-STABL

The goodness measure used for STABL, which is described in the HIDECS 3 report, is interesting because it signals a change in basic approach; it's called HIDECS3-STABL.

We start with the current, provisional partition of a set *M* into disjoint sets: $\{S_1, S_2, \ldots\}$. Then we define EXP as shown in Figure 12. The idea is to maximize EXP.

**The Intuition**: This measure is subtle. Note first that a goodness measure is not required to compute something meaningful, but only to reflect an idea of how to distinguish a better decomposition from a worse one. Both numerator and denominator are summed over sets in the decomposition, so the measure is looking at how good each set is, then combining that information.

Let's look at the numerator first. The lefthand side of the numerator can be rewritten as this:

$$\left(\frac{l_i}{l}\right)\frac{m(m-1)}{2}$$

$$m = \text{total number of variables} \tag{55}$$
$$l = \text{total number of links} \tag{56}$$
$$l_i = \text{total number of links totally within } S_i \tag{57}$$
$$s_i = \text{number of elements in } S_i \tag{58}$$
$$nom = \sum_i \left[ l_i \frac{m(m-1)}{2l} - \frac{s_i(s_i-1)+1}{2} \right] \tag{59}$$
$$denom = \sum_i \left( \frac{s_i(s_i-1)+1}{2} \right) 2^{-2s_i} \tag{60}$$
$$\text{EXP} = signum(nom) \frac{nom^2}{denom} \tag{61}$$

■ **Figure 12** HIDECS3-STABL

which is the product of two things: the fraction of links in set $s_i$ to the total number of links and the number of possible links—we saw the latter in earlier goodness measures (e.g., *nsq*1 in HIDECS2-Actual). The righthand side is the number of possible links completely within $s_i$, fudged a little. Let's call links that both start and terminate within a set *internal* (for set $s_i$ there are $l_i$ internal links), and those that start in a set and terminate outside it *crossovers*. Internal links go to cohesion and crossovers to coupling. First let's look at sets with only internal links. As $l_i$ increases from 0 to the largest it can be, $\frac{l_i}{l}$ increases linearly, as does the lefthand side of the numerator. The righthand side remains constant, so the numerator increases. This means that as the set grows more densely connected, its cohesion improves, and the measure increases.

Now suppose there are crossovers. Holding $l_i$ constant, as the number of crossovers (originating in $s_i$) increases from 0 to the maximum it can achieve, $l$ increases and therefore $\frac{l_i}{l}$ decreases, as does the numerator. This means that as the set acquires more crossovers, its coupling increases, causing the measure to decrease. The largest $l$ can get is the number of possible links, which means that using that as the ceiling for the lefthand side provides a sort of normalization for the numerator. Note that the righthand side of the numerator adds 1 to its numerator so that the righthand side is always positive. Overall, the numerator takes into account both cohesion and coupling, balancing them.

Now the denominator. When there are only two sets, the denominator has the shape of a frown as the size of the first set grows from 1 to its maximum; this means that the denominator causes the measure to favor balanced set sizes. When there are more than two, we see the same general behavior: favor balanced set sizes.

The numerator is squared because the denominator is a sort-of variance, as noted in the description of HIDECS2-Actual.

Stronger cohesion is how this measure increases, while coupling tends to decrease it; therefore one could say it is primarily for finding undiminished cohesion.

**Notes on "Notes on the Synthesis of Form"**

## E   Alexander Repudiates Tree Structure and Formal Design Methods

Christopher Alexander turned his back on two ideas presented in "Notes": first, that it is possible and desirable to attempt to partition a design "problem" into *disjoint* sets of concerns and a tree-like structure; second, that it is possible to devise a *method* separate from the practice of design to achieve such partitions and thus lay out a plan of design attack. Alexander's notion of a *pattern* is based on recognizing that absolute independence of concerns is not possible and that patterns can and should be developed piecemeal out of a designer's experience of designing real things.

### E.1  Against Tree Structure

Starting with "A City is Not a Tree," Alexander repeatedly turns his back on tree structure. To see how strong and persistent that repudiation is, I present a couple of quotes.

From "A City is Not a Tree":

> ***Whenever we have a tree structure, it means that within this structure no piece of any unit is ever connected to other units, except through the medium of that unit as a whole.***
>
> *The enormity of this restriction is difficult to grasp. It is a little as though the members of a family were not free to make friends outside the family, except when the family as a whole made a friendship.*
>
> *The structural simplicity of trees is like the compulsive desire for neatness and order that insists that the candlesticks on a mantelpiece be perfectly straight and perfectly symmetrical about the centre. The semilattice, by comparison, is the structure of a complex fabric; it is the structure of living things—of great paintings and symphonies.*
>
> *It must be emphasised, lest the orderly mind shrink in horror from anything that is not clearly articulated and categorised in tree form, that the ideas of overlap, ambiguity, multiplicity of aspect, and the semilattice, are not less orderly than the rigid tree, but more so. They represent a thicker, tougher, more subtle and more complex view of structure.* [14]

Also from "A City is Not a Tree":

> *The tree—though so neat and beautiful as a mental device, though it offers such a simple and clear way of dividing a complex entity into units—does not describe correctly the actual structure of naturally occurring cities, and does not describe the structure of the cities which we need.*
>
> *Now, why is it that so many designers have conceived cities as trees when the natural structure is in every case a semilattice? Have they done so deliberately, in the belief that a tree structure will serve the people of the city better? Or have they done it because they cannot help it, because they are trapped by a mental habit, perhaps even trapped by the way the mind works—because they cannot encompass the complexity of a semilattice in any convenient mental form, because*

> *the mind has an overwhelming predisposition to see trees wherever it looks and cannot escape the tree conception?*
>
> *I shall try to convince you that it is for this second reason that trees are being proposed and built as cities—that is, because designers, limited as they must be by the capacity of the mind to form intuitively accessible structures, cannot achieve the complexity of the semilattice in a single mental act.* [14]

Finally from "Battle for the Life and Beauty of the Earth: A Struggle Between Two World-Systems," Alexander's last book:

> *The principal features of a complex configuration are always created by overlap. Although this overlap may seem trivial, when we examine the overall design of* [a] *Persian carpet, you will see that this kind of overlap, and ambiguity, is essential and pervasive…. This is the glue in any system of wholes. Wholeness itself is directly created by this apparent overlap, or ambiguity. The greater the number of overlapping wholes, the more tightly bound the configuration is, and the more deeply the wholeness of the object shows itself to be.* [17]

### E.2  Against Formal Design Methods

In the preface to the paperback edition of "Notes" in 1971, Alexander apologizes for helping create a field of design science separate from design practice:

> *At the time I wrote this book, I was very much concerned with the formal definition of "independence," and the idea of using a mathematical method to discover systems of forces and diagrams which are independent.* But once the book was written, I discovered that it is quite unnecessary to use such a complicated and formal way of getting at the independent diagrams.
>
> *If you understand the need to create independent diagrams, which resolve, or solve, systems of interacting human forces, you will find that you can create, and develop, these diagrams piecemeal, one at a time, in the most natural way, out of your experience of buildings and design, simply by thinking about the forces which occur there and the conflicts between these forces.*
>
> *I have written about this realization and its consequences, in other, more recent works. But I feel it is important to say it also here, to make you alive to it before you read the book, since so many readers have focused on the* method which leads to *the creation of the diagrams, not on the* diagrams themselves*, and have even made a cult of following this method.*
>
> *Indeed, since the book was published, a whole academic field has grown up around the idea of "design methods"—and I have been hailed as one of the leading exponents of these so-called design methods. I am very sorry that this has happened, and want to state, publicly, that I reject the whole idea of design methods as a subject of study, since I think it is absurd to separate the study of designing from the practice of design.*

## F  Those Pesky Errors

Consider the matter of those 50 errors in the Indian Village interactions table. The errors are listed in Appendix B. At first I was shocked to see so many errors; I routinely found similar errors in several of the papers Alexander wrote in the early 1960s. But I was less shocked when I saw that the input of interactions to his programs took the form of putting 1s at row/col positions on punch cards, so that accidentally putting a 1 somewhere or leaving one out was not outrageous. You can see an image of one such input card in Appendix M.

Here is what Alexander wrote in the HIDECS 2 report about his approach to handling such errors:

> *The links of a graph, as used in this program, are non-directional. The matrix representing the links must therefore be symmetrical about its main diagonal, i.e. $m_{ij} = m_{ji}$. Also since no vertex may be linked to itself, the elements along the main diagonal, $m_{ii}$, must all be zero.*
>
> *The matrix is input from punched cards. Since a single error would make the matrix symmetrical [sic],[24] or introduce 1's on the main diagonal, and would thus destroy the conditions necessary for correct operation of the algorithms, the subprogram SYMET is used to check the input and remove errors of the above types. SYMET eradicates 1's on the main diagonal, and replaces both $m_{ij}$ and $m_{ji}$ by $(m_{ij} \wedge m_{ji})$ for all i and j, thus leaving a 1 in them only if both are initially 1.*
>
> *It turns out, incidentally, that SYMET is even more useful than originally intended. For a large design problem of the type which generates our input, it is not only hard to ensure accuracy in punching, but also hard to decide just which point pairs are linked. If the decision for $m_{ij}$ is made independently of the decision for $m_{ji}$, it is almost impossible, in practice, to make all these decisions consistent with the formal symmetry required. With the program SYMET in operation, however, it is possible to generate the data at the card punch, without handchecking it, with the assurance that it will be machine checked and that only the "most certain" pairs—those where a link has been defined for both $m_{ij}$ and $m_{ji}$—will be treated as linked.*

This description demonstrates a couple of Alexander's common shortcomings in his early papers: simple (typically trivial) errors and overlooking opportunities.[25] The simple error is that in the phrase "…a single error would make the matrix symmetrical,…" "symmetrical" should be either "asymmetrical" or "nonsymmetrical." For a reader puzzling out the complexities of the program, this trivial error can cause perplexity. The overlooked opportunity is that instead of celebrating how (more) useful SYMET is by machine checking for and deleting uncertain pairs that might have been produced by people at the card punch, Alexander could have noticed that by programming the computer to point out these inconsistencies—by outputting warnings—SYMET could helpfully work with the programmer to find and correct mistakes correctly. Hand checking would be dramatically simplified.

---

[24] Should be "asymmetrical."

[25] While doing a close reading of an earlier paper of his—"A Result in Visual Aesthetics" [2]— I noticed a handful of similarly erroneous statements and data [21].

The Python version of HIDECS 2 confirms Alexander's description, as does the flow chart for SYMET.

The fact that the HIDECS 2 program included the subprogram SYMET means, furthermore, that Alexander was aware of the possibility of errors, but the fact that he did not correct them in "Notes" or in the interactions for the design problem he explored in "Community and Privacy" [20] (discussed in Appendix Q) tells me he was sometimes not obsessive about details—at least not in his early papers. Contrast that with what he said about fine details in a building in 1993 in "A Foreshadowing of 21$^{st}$ Century Art" [9]:

> *...if we hope to make buildings in which the rooms and building feel harmonious— we too, must make sure that the structure is correct down to $1/8^{th}$ of an inch. Any structure which is more gross, and which leaves this last eighth of an inch, rough, or uncalculated, or inharmonious—will inevitably be crude.*

This leaves one mystery: where did these 50 errors come from? They appear in "Notes" as text (reproduced in Appendix B). Presumably the interactions were determined by hand somehow by Alexander. The text of them in "Notes" could have been derived directly from that set, which means that the errors were in the hand-created set; then these errors were propagated—perhaps accurately—to the punch cards, and were subsequently "corrected" by SYMET. Or the hand-created set contained no errors, but errors were introduced into the punch cards, and the text in "Notes" was derived from those cards—by a program that simply printed out the matrix of interactions. I could not find the hand-created set of interactions, but there is a subroutine called "PTMAT" in the HIDECS 2 code that prints out the matrix of interactions after SYMET has operated on it. I found no evidence the text of interactions in "Notes" came from PTMAT.

I don't know the protocols at the MIT Computation Center in the early 1960s for using the IBM 709[26] Alexander and Manheim used for their programming, but fifteen years later at Stanford in similar circumstances, one submitted a card deck to an operator who would feed the punch cards into the machine, returning a printout of the results some time later—in some cases, the next day. In such an environment, it was common to deal with errors and mistakes with as few such interactions as possible, and that could include trying to build in error recovery mechanisms like SYMET.

Alexander's shortcomings, however, don't diminish the strength of his research, thinking, and writing, which many believe (me included) essential to 20$^{th}$ and 21$^{st}$ century design thinking.

---

[26] An IBM 7090 was also used for some of their runs; it was located at the Smithsonian Astrophysical Observatory in Cambridge, Massachusetts.

### G  A Telltale Anomaly, Details

I wanted to see how well Alexander's program did partitioning the Entire Village—the hardest partition of all. Alexander presents a partition of the whole problem into four sets, A, B, C, D. As noted, his program actually produces a binary partition of the whole problem $(X, Y)$, and then each of those was further partitioned into two, yielding four. There are two possible approaches for how this could yield the four Alexander shows: the balanced approach is that the Entire Village is partitioned into two sets, $X$ and $Y$, which are each partitioned into the presented four with $X = A \cup B$, $X = A \cup C$, or $X = A \cup D$?[27] as possibilities. The imbalanced approach is that the Entire Village is partitioned, for example, into A and $X = B \cup C \cup D$, followed by, for example, $X$ being partitioned into B and $Y = C \cup D$, followed by $Y$ being partitioned into C and D; there are twelve such imbalanced possibilities. I started with the balanced possibilities—which I believe is what Alexander did—and found the anomaly; I did not explore the imbalanced possibilities.

Here are the balanced possibilities:

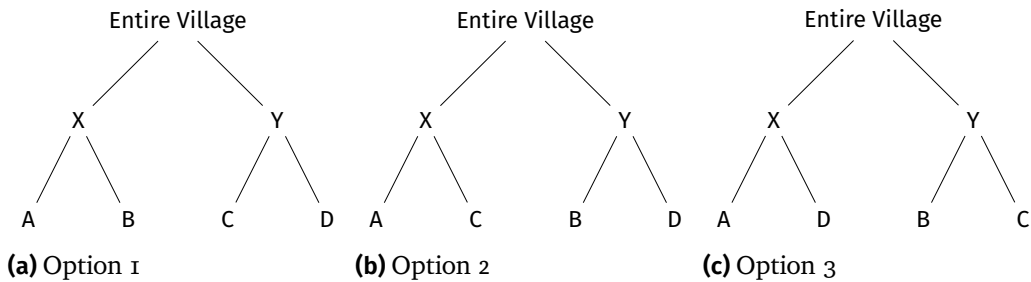

**(a)** Option 1   **(b)** Option 2   **(c)** Option 3

We know what vertices are in A because we know what vertices are in A1, A2, and A3: they are listed on page 151 of "Notes":

| Group | Elements |
|---|---|
| A1 | 7, 53, 57, 59, 60, 72, 125, 126, 128 |
| A2 | 31, 34, 36, 52, 54, 80, 94, 106, 136 |
| A3 | 37, 38, 50, 55, 77, 91, 103 |

Similarly for B, C, and D. Therefore we know what vertices would be in $X$ if $X = A \cup B$ and in $Y$ if $Y = C \cup D$, for example.

To find out which two came from the same initial partition, I tried all possible pairings—that is, I tried Options 1, 2, and 3—and the pairing that produced the best goodness for $X$ and $Y$ using the goodness measure is Option 1. For concreteness, here are the raw values (smaller numbers are better, so $-645$ is better than $-562$):

| Option | Goodness |
|---|---|
| Option 1 | $-645.04$ |
| Option 2 | $-434.40$ |
| Option 3 | $-562.65$ |

---

[27] Note that if $X = A \cup B$, then $Y = C \cup D$, etc.

I guessed Option 1 was what Alexander's program did. Then I tried running my version of HIDECS 2 on the Entire Village; its result at the first level, $X_1$ and $Y_1$, measured out to $-655.12$—clearly better than all the options derived from Alexander's partition into four sets. I expected that if my program took that $X_1$, it would produce $A_1$ and $B_1$ that would measure out better than Alexander's A and B; and taking that $Y_1$, it would produce $C_1$ and $D_1$ that would also measure out better. This was naïve: the resulting partitions from my program were not much like Alexander's; it proved problematic to come up with an apples / apples comparison.

While trying to figure out how to proceed, I ran an exhaustive pairwise computation of the goodness measure on Alexander's A, B, C, and D:

| Pairs | Goodness |
|---|---|
| A & B: | $-197.83$ |
| A & C: | $-257.00$ |
| A & D: | $-197.98$ |
| B & C: | $-341.70$ |
| B & D: | $-345.84$ |
| C & D: | $-297.75$ |

From this table I guessed that Alexander's program partitioned the Entire Village into $X = A \cup C$ and $Y = B \cup D$. This is the worst of the three options. When I used those for starting points and derived my versions of $A_2$, $B_2$, $C_2$, and $D_2$, they were exactly the same as Alexander's.

Stated bluntly: the overall best partition (for A, B, C, D) is not necessarily obtained by doing the best job starting at the top and working down to get the best $X$ and $Y$, followed by getting the best A & B from $X$ and the best C & D from $Y$. This is what Alexander meant in the first of his three observed weaknesses of HIDECS 2 as discussed in the HIDECS 3 report: "the holistic relatedness of system and subsystems is not properly taken into account." His demonstration of this is in Appendix H.

## H  Alexander's Demonstration Against Top-Down

In the HIDECS 3 report Alexander presents a clear demonstration why his top-down (binary) approach to decomposition in HIDECS 2 cannot work. It starts with an elegant decomposition example (Figure 14).

Suppose that the first-level partition of a set of nodes is into the sets $L = L_0 \cup L_1 \cup L_2 \cup L_3$ and $R = R_0 \cup R_1 \cup R_2 \cup R_3$, and that the node $x$ has yet to be assigned. Further suppose that eventually $L$ will be decomposed into the sets $\{L_0, L_1, L_2, L_3\}$ and $R$ into $\{R_0, R_1, R_2, R_3\}$. Node $x$ has three links to $L_2$ and one each to $R_0, R_1, R_2, R_3$. At that first level, when there are only $L$ and $R$, $x$ has three links to $L$ and four links to $R$, and therefore $x$ will be placed in $R$. When $R$ is decomposed further, $x$ will join one of the $R_i$, with which it has only one link. The node $x$ belongs in $L_2$. Without being able to look ahead $x$ will be misplaced. Returning to the family analogy, someone with many friends in another family might be considered a member of that other family and not of their real family.

**Notes on "Notes on the Synthesis of Form"**

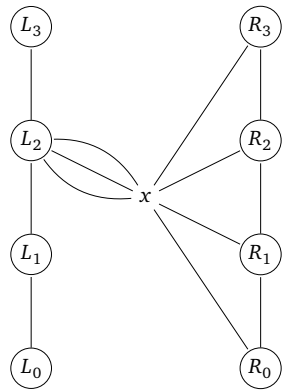

**Figure 14** Alexander's Elegant Example

This led Alexander to a more bottom-up approach to decomposition.

## I Comparing Decomposiitons

I compared Alexander's decomposition to ones my programs did: the first was to produce the decomposition down to level 4 (Alexander's lowest level) using HIDECS2-Actual, and the second using HIDECS2-rpg. My best attempt at re-creating a possible such tree is in Figure 15—the likely partitions $X$ and $Y$ are filled in.

### I.1 Using My Recoding of HIDECS 2

Let's take a look at what my program searching about 50 times more starting sets does for the Indian Village: Figure 16.

Notice there are 16 leaf groups at level 4. Let's compare this to Alexander's decomposition. First some definitions for partitions in his decomposition:

$$CA(\pi_1) = \{A \cup C, B \cup D\} \tag{62}$$
$$CA(\pi_2) = \{A, B, C, D\} \tag{63}$$
$$CA(\pi_4) = \{A1, A2, A3, B1, B2, B3, B4, C1, C2, D1, D2, D3\} \tag{64}$$

Now for my decomposition:

$$Lf(node) = \text{union of all leaves reachable from } node \tag{65}$$
$$Leaves(n) = \{Lf(node) \mid Label(node) = n\} \tag{66}$$
$$\text{rpg}(\pi_1) = Leaves(1) \tag{67}$$
$$\text{rpg}(\pi_2) = Leaves(2) \tag{68}$$
$$\text{rpg}(\pi_4) = Leaves(4) \tag{69}$$

Simply stated: for example, to figure out rpg($\pi_2$), find all the circled nodes labeled 2 and collect the sets of leaves that are under them—don't union them, just collect them.

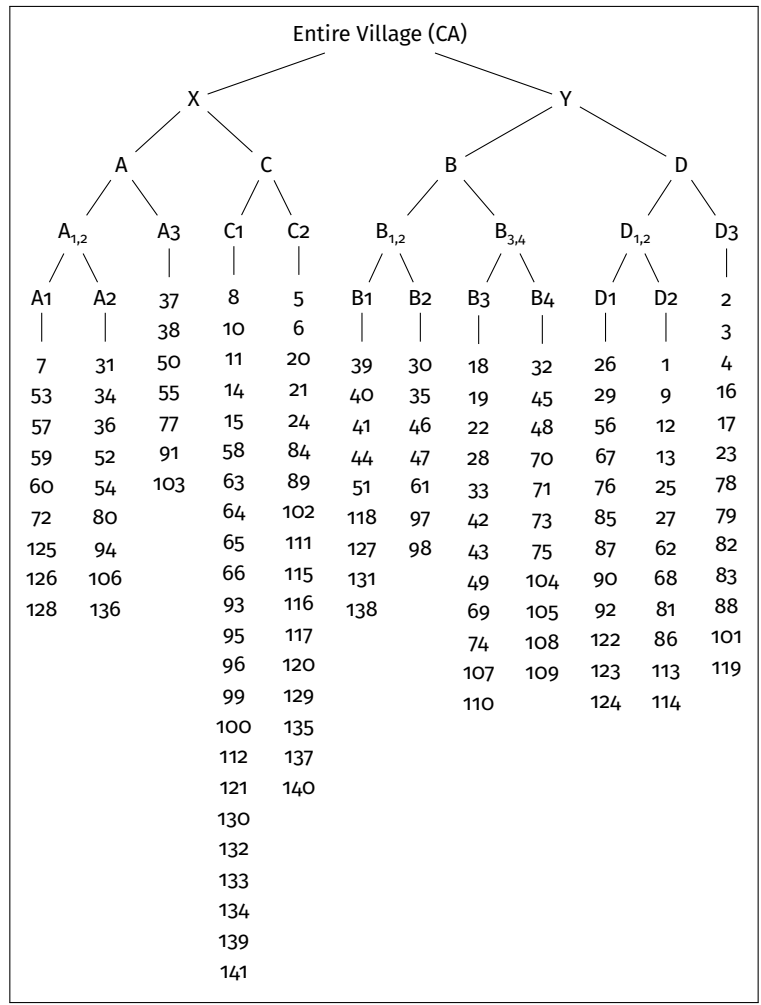

**Figure 15** Entire Village (CA)

The partition rpg($\pi_4$) has 16 sets. We are going to look at levels in the tree, which represent partitions of the entire Indian Village into different granularities of partitions.

I'll use HIDECS2-Notes, otherwise known as $R(\pi)$, as the goodness measure. Note that in the column labeled **CA**, the values shown are for $R(CA(\pi))$, and in the column labeled **rpg**$_\pi$, the values shown are for $R(rpg(\pi))$ (green indicates better result):

| Partition ($\pi$) | CA | rpg$_\pi$ |
|---|---|---|
| $\pi_1$ | −434.40 | −655.12 |
| $\pi_2$ | −945.57 | −940.48 |
| $\pi_4$ | −1072.62 | −1182.20 |

**Notes on "Notes on the Synthesis of Form"**

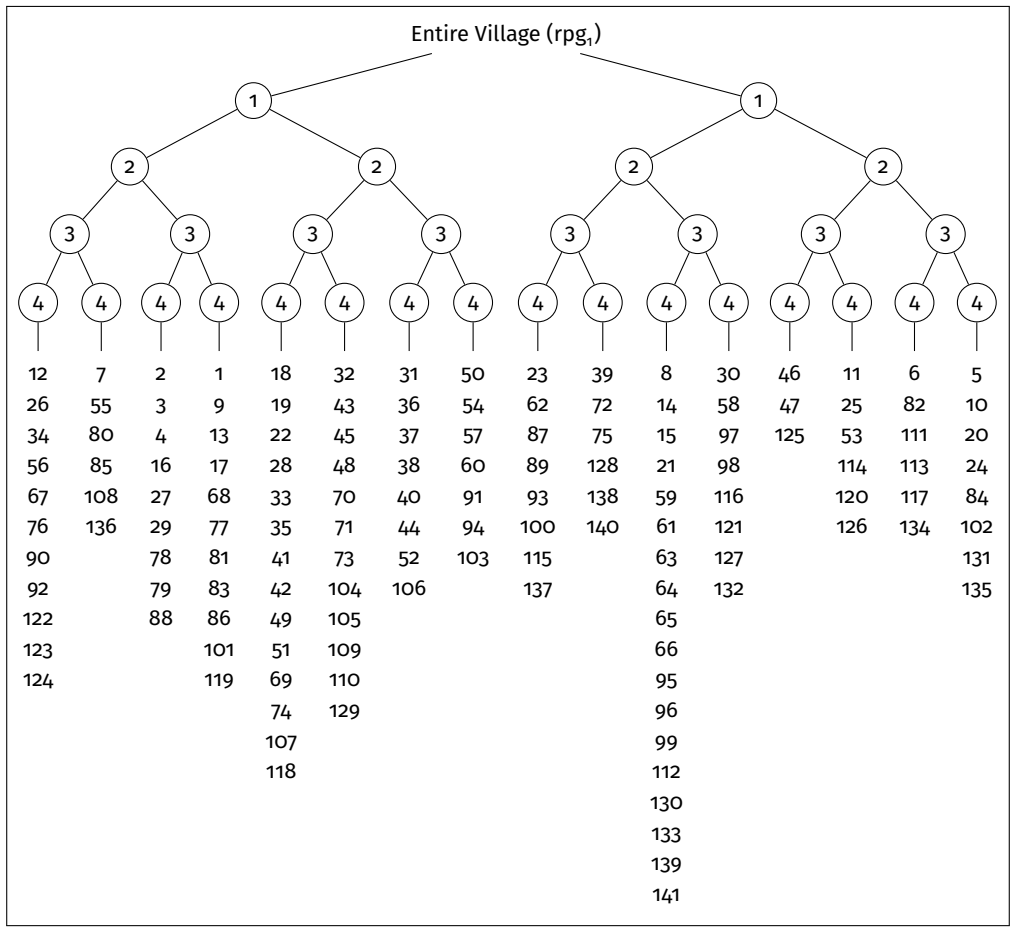

◼ **Figure 16**   Entire Village (rpg$_1$)

### I.2  Using My Dreamed Up Goodness Measure

I did another experiment where I dreamed up my own cohesion / coupling goodness measure: HIDECS2-rpg, described in Appendix D. It produced the decomposition shown in Figure 17

In a manner similar to the above comparison of decompositions, this one measures out like this:

| Partition ($\pi$) | CA | rpg$_1$ |
|---|---|---|
| $\pi_1$ | −434.40 | −652.82 |
| $\pi_2$ | −945.57 | −891.87 |
| $\pi_4$ | −1072.62 | −1089.22 |

## J  Requirements Text Comparison

Here are comparisons of Alexander's decomposition of the Indian Village to decompositions two of my programs did: the first is simply my version of Alexander's HIDECS 2

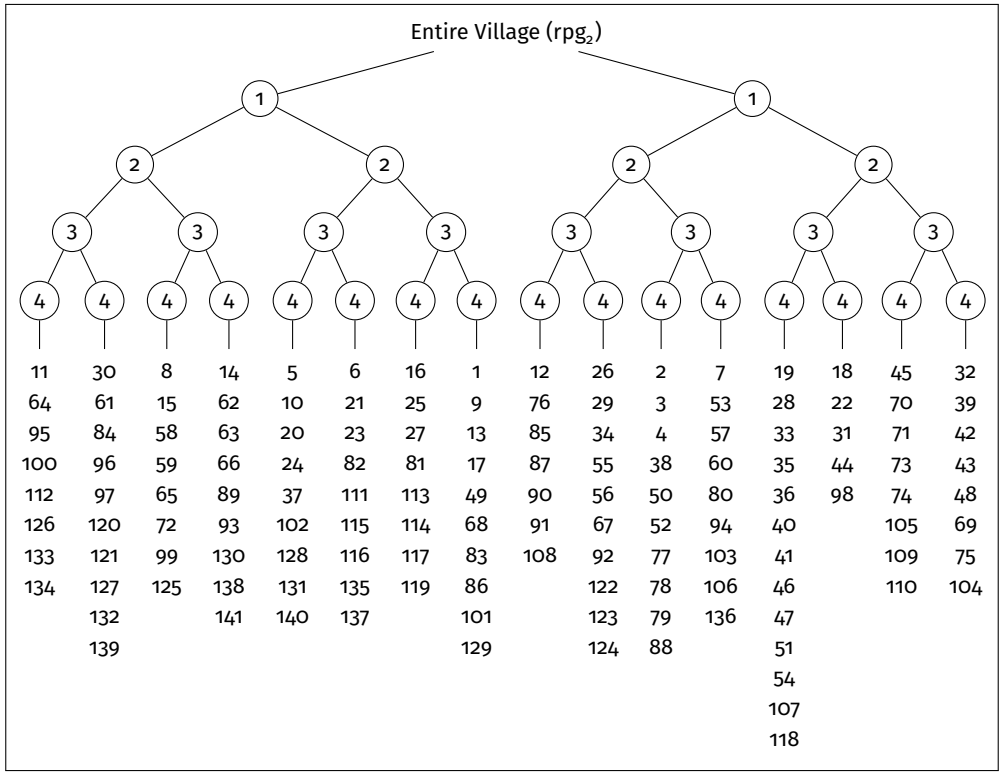

■ **Figure 17** Entire Village (rpg$_2$)

program but searching many more starting partitions, and the second is the same program but using my goodness measure, HIDECS2-rpg. You can use these tables to judge how good the decompositions are—by examining them for coherence. While doing this, keep in mind what Alexander wrote about the partition:

> *We must remember that the hierarchy of sets which the tree defines will not always be easy to understand. Even in some of the smaller sets which contain only half a dozen variables, these variables will often seem disparate, and their juxtaposition may be startling. The relevance of each variable is only to be properly understood after careful examination of its functional relation to the other variables in the set. Since the potential coherence of such a set of variables comes from its physical implications, it can only be grasped graphically, by means of a constructive diagram that brings out these implications.* [5]

The first corresponds to the tree labeled *Entire Village (rpg$_1$)* (Figure 16); it uses HIDECS2-Decomp and examines 50 times more initial partitions. The following tables (with light green backgrounds) show the text for the best pairings of partitions from Alexander and my program. Note that my program produces four more sets than Alexander reports—these are the "Unpaired rpg Sets."

## Notes on "Notes on the Synthesis of Form"

| A1 (2 in common) | | rpg Set 1 (paired with A1) | |
|---|---|---|---|
| 53 | Upgrading of cattle. | 53 | Upgrading of cattle. |
| 126 | Close contact with village-level worker. | 126 | Close contact with village-level worker. |
| 7 | Cattle treated as sacred, and vegetarian attitude. | 11 | Marriage is to person from another village. |
| 57 | Protection of cattle from disease. | 25 | Assistance for physically handicapped, aged, widows. |
| 59 | Efficient use and marketing of dairy products. | 114 | Development of women's independent activities. |
| 60 | Minimize the use of animal traction to take pressure off shortage. | 120 | Curative measures for disease available to villagers. |
| 72 | Prevent famine if monsoon fails. | | |
| 125 | Prevent malnutrition. | | |
| 128 | Price assurance for crops. | | |

| A2 (4 in common) | | rpg Set 2 (paired with A2) | |
|---|---|---|---|
| 31 | Efficient distribution of fertilizer, manure, seed, from village storage to fields. | 31 | Efficient distribution of fertilizer, manure, seed, from village storage to fields. |
| 36 | Protection of crops from thieves, cattle, goats, monkeys, etc. | 36 | Protection of crops from thieves, cattle, goats, monkeys, etc. |
| 52 | Improve quantity of fodder available. | 52 | Improve quantity of fodder available. |
| 106 | Young trees need protection from goats, etc. | 106 | Young trees need protection from goats, etc. |
| 34 | Full collection of natural manure (animal and human). | 37 | Provision of storage for distributing and marketing crops. |
| 54 | Provision for feeding cattle. | 38 | Provision of threshing floor and its protection from marauders. |
| 80 | Security for cattle. | 40 | Best food grain crop. |
| 94 | Provision for animal traffic. | 44 | Crops must be brought home from fields. |
| 136 | Accommodation of wandering caste groups, incoming labor, etc. | | |

| A3 (3 in common) | | rpg Set 3 (paired with A3) | |
|---|---|---|---|
| 50 | Protected storage of fodder. | 50 | Protected storage of fodder. |
| 91 | Provision and storage of fuel. | 91 | Provision and storage of fuel. |
| 103 | Bullock cart access to house for bulk of grain, fodder. | 103 | Bullock cart access to house for bulk of grain, fodder. |
| 37 | Provision of storage for distributing and marketing crops. | 54 | Provision for feeding cattle. |
| 38 | Provision of threshing floor and its protection from marauders. | 57 | Protection of cattle from disease. |
| 55 | Cattle access to water. | 60 | Minimize the use of animal traction to take pressure off shortage. |
| 77 | Village and individual houses must be protected from fire. | 94 | Provision for animal traffic. |

| B1 (2 in common) | | rpg Set 4 (paired with B1) | |
|---|---|---|---|
| 39 | Best cotton and cash crop. | 39 | Best cotton and cash crop. |
| 138 | Achieve economic independence so as not to strain national transportation and resources. | 138 | Achieve economic independence so as not to strain national transportation and resources. |
| 40 | Best food grain crop. | 72 | Prevent famine if monsoon fails. |
| 41 | Good vegetable crop. | 75 | Drainage of land to prevent waterlogging, etc. |
| 44 | Crops must be brought home from fields. | 128 | Price assurance for crops. |
| 51 | Improve quality of fodder available. | 140 | Develop rural community spirit: destroy selfishness, isolationism. |
| 118 | Demonstration projects which spread by example. | | |
| 127 | Contact with block development officer and extension officers. | | |
| 131 | Panchayat must have more power and respect. | | |

| b2 (3 in common) | | rpg Set 5 (paired with b2) | |
|---|---|---|---|
| 30 | Efficient and rapid distribution of seeds, fertilizer, etc., from block HQ. | 30 | Efficient and rapid distribution of seeds, fertilizer, etc., from block HQ. |
| 97 | Minimize transportation costs for bulk produce (grain, potatoes, etc.). | 97 | Minimize transportation costs for bulk produce (grain, potatoes, etc.). |
| 98 | Daily produce requires cheap and constant (monsoon) access to market. | 98 | Daily produce requires cheap and constant (monsoon) access to market. |
| 35 | Protection of crops from insects, weeds, disease. | 58 | Development of other animal industry. |
| 46 | Respect for traditional agricultural practices. | 116 | Improvement of adult literacy. |
| 47 | Need for new implements when old ones are damaged, etc. | 121 | Facilities for birth, pre- and post-natal care, birth control. |
| 61 | Sufficient fluid employment for laborers temporarily (seasonally) out of work. | 127 | Contact with block development officer and extension officers. |
| | | 132 | Need to develop projects which benefit from government subsidies. |

| b3 (10 in common) | | rpg Set 6 (paired with b3) | |
|---|---|---|---|
| 18 | Need to divide land among sons of successive generations. | 18 | Need to divide land among sons of successive generations. |
| 19 | People want to own land personally. | 19 | People want to own land personally. |
| 22 | Abolition of Zamindari and uneven land distribution. | 22 | Abolition of Zamindari and uneven land distribution. |
| 28 | Proper boundaries of ownership and maintenance responsibility. | 28 | Proper boundaries of ownership and maintenance responsibility. |
| 33 | Fertile land to be used to best advantage. | 33 | Fertile land to be used to best advantage. |
| 42 | Efficient plowing, weeding, harvesting, leveling. | 42 | Efficient plowing, weeding, harvesting, leveling. |
| 49 | Cooperative farming. | 49 | Cooperative farming. |
| 69 | Fullest possible irrigation benefit derived from available water. | 69 | Fullest possible irrigation benefit derived from available water. |
| 74 | Maintenance of irrigation facilities. | 74 | Maintenance of irrigation facilities. |
| 107 | Soil conservation. | 107 | Soil conservation. |
| 43 | Consolidation of land. | 35 | Protection of crops from insects, weeds, disease. |
| 110 | Prevent land erosion. | 41 | Good vegetable crop. |
| | | 51 | Improve quality of fodder available. |
| | | 118 | Demonstration projects which spread by example. |

| b4 (9 in common) | | rpg Set 7 (paired with b4) | |
|---|---|---|---|
| 32 | Reclamation and use of uncultivated land. | 32 | Reclamation and use of uncultivated land. |
| 45 | Development of horticulture. | 45 | Development of horticulture. |
| 48 | Scarcity of land. | 48 | Scarcity of land. |
| 70 | Full collection of underground water for irrigation. | 70 | Full collection of underground water for irrigation. |
| 71 | Full collection of monsoon water for use. | 71 | Full collection of monsoon water for use. |
| 73 | Conservation of water resources for future. | 73 | Conservation of water resources for future. |
| 104 | Plant ecology to be kept healthy. | 104 | Plant ecology to be kept healthy. |
| 105 | Insufficient forest land. | 105 | Insufficient forest land. |
| 109 | Reclamation of eroded land, gullies, etc. | 109 | Reclamation of eroded land, gullies, etc. |
| 75 | Drainage of land to prevent waterlogging, etc. | 43 | Consolidation of land. |
| 108 | Road and dwelling erosion. | 110 | Prevent land erosion. |
| | | 129 | Factions refuse to cooperate or agree. |

**Notes on "Notes on the Synthesis of Form"**

| C1 (15 in common) | | rpg Set 8 (paired with C1) | |
|---|---|---|---|
| 8 | Members of castes maintain their caste profession as far as possible. | 8 | Members of castes maintain their caste profession as far as possible. |
| 14 | Economic integration of village on payment-in-kind basis. | 14 | Economic integration of village on payment-in-kind basis. |
| 15 | Modern move toward payment in cash. | 15 | Modern move toward payment in cash. |
| 63 | Development of village industry. | 63 | Development of village industry. |
| 64 | Simplify the mobility of labor, to and from villages, and to and from fields and industries and houses. | 64 | Simplify the mobility of labor, to and from villages, and to and from fields and industries and houses. |
| 65 | Diversification of villages' economic base—not all occupations agricultural. | 65 | Diversification of villages' economic base—not all occupations agricultural. |
| 66 | Efficient provision and use of power. | 66 | Efficient provision and use of power. |
| 95 | Access to bus as near as possible. | 95 | Access to bus as near as possible. |
| 96 | Access to railway station. | 96 | Access to railway station. |
| 99 | Industry requires strong transportation support. | 99 | Industry requires strong transportation support. |
| 112 | Access to a secondary school. | 112 | Access to a secondary school. |
| 130 | Need for increased incentives and aspirations. | 130 | Need for increased incentives and aspirations. |
| 133 | Social integration with neighboring villages. | 133 | Social integration with neighboring villages. |
| 139 | Proper connection with bridges, roads, hospitals, schools, | 139 | Proper connection with bridges, roads, hospitals, schools, |
| 141 | Prevent migration of young people and harijans to cities. | 141 | Prevent migration of young people and harijans to cities. |
| 10 | Need for elaborate weddings. | 21 | Eradication of untouchability. |
| 11 | Marriage is to person from another village. | 59 | Efficient use and marketing of dairy products. |
| 58 | Development of other animal industry. | 61 | Sufficient fluid employment for laborers temporarily (seasonally) out of work. |
| 93 | Lighting. | | |
| 100 | Provision for bicycle age in every village by 1965. | | |
| 121 | Facilities for birth, pre- and post-natal care, birth control. | | |
| 132 | Need to develop projects which benefit from government subsidies. | | |
| 134 | Wish to keep up with achievements of neighboring villages. | | |

| C2 (6 in common) | | rpg Set 9 (paired with C2) | |
|---|---|---|---|
| 5 | Provision for festivals and religious meetings. | 5 | Provision for festivals and religious meetings. |
| 20 | People of different factions prefer to have no contact. | 20 | People of different factions prefer to have no contact. |
| 24 | Place for village events—dancing, plays, singing, etc., wrestling. | 24 | Place for village events—dancing, plays, singing, etc., wrestling. |
| 84 | Accommodation for panchayat records, meetings, etc. | 84 | Accommodation for panchayat records, meetings, etc. |
| 102 | Accommodation for processions. | 102 | Accommodation for processions. |
| 135 | Spread of official information about taxes, elections, etc. | 135 | Spread of official information about taxes, elections, etc. |
| 6 | Wish for temples. | 10 | Need for elaborate weddings. |
| 21 | Eradication of untouchability. | 131 | Panchayat must have more power and respect. |
| 89 | Provision of goods, for sale. | | |
| 111 | Provision for primary education. | | |
| 115 | Opportunity for youth activities. | | |
| 116 | Improvement of adult literacy. | | |
| 117 | Spread of information about birth control, disease, etc. | | |
| 120 | Curative measures for disease available to villagers. | | |
| 129 | Factions refuse to cooperate or agree. | | |
| 137 | Radio communication. | | |
| 140 | Develop rural community spirit: destroy selfishness, isolationism. | | |

| D1 (9 in common) | | rpg Set 10 (paired with D1) | |
|---|---|---|---|
| 26 | Sentimental system: wish not to destroy old way of life; love of present habits governing bathing, food, etc. | 26 | Sentimental system: wish not to destroy old way of life; love of present habits governing bathing, food, etc. |
| 56 | Sheltered accommodation for cattle (sleeping, milking, feeding). | 56 | Sheltered accommodation for cattle (sleeping, milking, feeding). |
| 67 | Drinking water to be good, sweet. | 67 | Drinking water to be good, sweet. |
| 76 | Flood control to protect houses, roads, etc. | 76 | Flood control to protect houses, roads, etc. |
| 90 | Better provision for preparing meals. | 90 | Better provision for preparing meals. |
| 92 | House has to be cleaned, washed, drained. | 92 | House has to be cleaned, washed, drained. |
| 122 | Disposal of human excreta. | 122 | Disposal of human excreta. |
| 123 | Prevent breeding germs and disease starters. | 123 | Prevent breeding germs and disease starters. |
| 124 | Prevent spread of human disease by carriers, infection, contagion. | 124 | Prevent spread of human disease by carriers, infection, contagion. |
| 29 | Provision for daily bath, segregated by sex, caste, and age. | 12 | Extended family is in one house. |
| 85 | Everyone's accommodation for sitting and sleeping should be protected from rain. | 34 | Full collection of natural manure (animal and human). |
| 87 | Safe storage of goods. | | |

# Notes on "Notes on the Synthesis of Form"

| **D2** (6 in common) | | **rpg Set 11** (paired with D2) | |
|---|---|---|---|
| 1 | Harijans regarded as ritually impure, untouchable, etc. | 1 | Harijans regarded as ritually impure, untouchable, etc. |
| 9 | Members of one caste like to be together and separate from others, and will not eat or drink together. | 9 | Members of one caste like to be together and separate from others, and will not eat or drink together. |
| 13 | Family solidarity and neighborliness even after separation. | 13 | Family solidarity and neighborliness even after separation. |
| 68 | Easy access to drinking water. | 68 | Easy access to drinking water. |
| 81 | Security for women and children. | 81 | Security for women and children. |
| 86 | No overcrowding. | 86 | No overcrowding. |
| 12 | Extended family is in one house. | 17 | Village has fixed men's social groups. |
| 25 | Assistance for physically handicapped, aged, widows. | 77 | Village and individual houses must be protected from fire. |
| 27 | Family is authoritarian. | 83 | In summer people sleep in open. |
| 62 | Provision of cottage industry and artisan workshops and training. | 101 | Pedestrian traffic within village. |
| 113 | Good attendance in school. | 119 | Efficient use of school; no distraction of students. |
| 114 | Development of women's independent activities. | | |

| **D3** (7 in common) | | **rpg Set 12** (paired with D3) | |
|---|---|---|---|
| 2 | Proper disposal of dead. | 2 | Proper disposal of dead. |
| 3 | Rules about house door not facing south. | 3 | Rules about house door not facing south. |
| 4 | Certain water and certain trees are thought of as sacred. | 4 | Certain water and certain trees are thought of as sacred. |
| 16 | Women gossip extensively while bathing, fetching water, on way to field latrines, etc. | 16 | Women gossip extensively while bathing, fetching water, on way to field latrines, etc. |
| 78 | Shade for sitting and walking. | 78 | Shade for sitting and walking. |
| 79 | Provision of cool breeze. | 79 | Provision of cool breeze. |
| 88 | Place to wash and dry clothes. | 88 | Place to wash and dry clothes. |
| 17 | Village has fixed men's social groups. | 27 | Family is authoritarian. |
| 23 | Men's groups chatting, smoking, even late at night. | 29 | Provision for daily bath, segregated by sex, caste, and age. |
| 82 | Provision for children to play (under supervision). | | |
| 83 | In summer people sleep in open. | | |
| 101 | Pedestrian traffic within village. | | |
| 119 | Efficient use of school; no distraction of students. | | |

| **Unpaired rpg Set 13** | |
|---|---|
| 23 | Men's groups chatting, smoking, even late at night. |
| 62 | Provision of cottage industry and artisan workshops and training. |
| 87 | Safe storage of goods. |
| 89 | Provision of goods, for sale. |
| 93 | Lighting. |
| 100 | Provision for bicycle age in every village by 1965. |
| 115 | Opportunity for youth activities. |
| 137 | Radio communication. |

| **Unpaired rpg Set 14** | |
|---|---|
| 46 | Respect for traditional agricultural practices. |
| 47 | Need for new implements when old ones are damaged, etc. |
| 125 | Prevent malnutrition. |

| Unpaired rpg Set 15 | |
|---|---|
| 6 | Wish for temples. |
| 82 | Provision for children to play (under supervision). |
| 111 | Provision for primary education. |
| 113 | Good attendance in school. |
| 117 | Spread of information about birth control, disease, etc. |
| 134 | Wish to keep up with achievements of neighboring villages. |

| Unpaired rpg Set 16 | |
|---|---|
| 7 | Cattle treated as sacred, and vegetarian attitude. |
| 55 | Cattle access to water. |
| 80 | Security for cattle. |
| 85 | Everyone's accommodation for sitting and sleeping should be protected from rain. |
| 108 | Road and dwelling erosion. |
| 136 | Accommodation of wandering caste groups, incoming labor, etc. |

The second corresponds to the tree labeled *Entire Village (rpg$_2$)* (Figure 17); it uses HIDECS2-rpg and examines 50 times more initial partitions. The following tables (with light red backgrounds) show the text for the best pairings of partitions from Alexander and my program. Note that my program produces four more sets than Alexander reports—these are the "Unpaired rpg Sets."

| A1 (3 in common) | | rpg Set 1 (paired with A1) | |
|---|---|---|---|
| 59 | Efficient use and marketing of dairy products. | 59 | Efficient use and marketing of dairy products. |
| 72 | Prevent famine if monsoon fails. | 72 | Prevent famine if monsoon fails. |
| 125 | Prevent malnutrition. | 125 | Prevent malnutrition. |
| 7 | Cattle treated as sacred, and vegetarian attitude. | 8 | Members of castes maintain their caste profession as far as possible. |
| 53 | Upgrading of cattle. | 15 | Modern move toward payment in cash. |
| 57 | Protection of cattle from disease. | 58 | Development of other animal industry. |
| 60 | Minimize the use of animal traction to take pressure off shortage. | 65 | Diversification of villages' economic base—not all occupations agricultural. |
| 126 | Close contact with village-level worker. | 99 | Industry requires strong transportation support. |
| 128 | Price assurance for crops. | | |

| A2 (4 in common) | | rpg Set 2 (paired with A2) | |
|---|---|---|---|
| 80 | Security for cattle. | 80 | Security for cattle. |
| 94 | Provision for animal traffic. | 94 | Provision for animal traffic. |
| 106 | Young trees need protection from goats, etc. | 106 | Young trees need protection from goats, etc. |
| 136 | Accommodation of wandering caste groups, incoming labor, etc. | 136 | Accommodation of wandering caste groups, incoming labor, etc. |
| 31 | Efficient distribution of fertilizer, manure, seed, from village storage to fields. | 7 | Cattle treated as sacred, and vegetarian attitude. |
| 34 | Full collection of natural manure (animal and human). | 53 | Upgrading of cattle. |
| 36 | Protection of crops from thieves, cattle, goats, monkeys, etc. | 57 | Protection of cattle from disease. |
| 52 | Improve quantity of fodder available. | 60 | Minimize the use of animal traction to take pressure off shortage. |
| 54 | Provision for feeding cattle. | 103 | Bullock cart access to house for bulk of grain, fodder. |

# Notes on "Notes on the Synthesis of Form"

| A3 (1 in common) | | rpg Set 3 (paired with A3) | |
|---|---|---|---|
| 91 | Provision and storage of fuel. | 91 | Provision and storage of fuel. |
| 37 | Provision of storage for distributing and marketing crops. | 12 | Extended family is in one house. |
| 38 | Provision of threshing floor and its protection from marauders. | 76 | Flood control to protect houses, roads, etc. |
| 50 | Protected storage of fodder. | 85 | Everyone's accommodation for sitting and sleeping should be protected from rain. |
| 55 | Cattle access to water. | 87 | Safe storage of goods. |
| 77 | Village and individual houses must be protected from fire. | 90 | Better provision for preparing meals. |
| 103 | Bullock cart access to house for bulk of grain, fodder. | 108 | Road and dwelling erosion. |

| B1 (4 in common) | | rpg Set 4 (paired with B1) | |
|---|---|---|---|
| 40 | Best food grain crop. | 40 | Best food grain crop. |
| 41 | Good vegetable crop. | 41 | Good vegetable crop. |
| 51 | Improve quality of fodder available. | 51 | Improve quality of fodder available. |
| 118 | Demonstration projects which spread by example. | 118 | Demonstration projects which spread by example. |
| 39 | Best cotton and cash crop. | 19 | People want to own land personally. |
| 44 | Crops must be brought home from fields. | 28 | Proper boundaries of ownership and maintenance responsibility. |
| 127 | Contact with block development officer and extension officers. | 33 | Fertile land to be used to best advantage. |
| 131 | Panchayat must have more power and respect. | 35 | Protection of crops from insects, weeds, disease. |
| 138 | Achieve economic independence so as not to strain national transportation and resources. | 36 | Protection of crops from thieves, cattle, goats, monkeys, etc. |
| | | 46 | Respect for traditional agricultural practices. |
| | | 47 | Need for new implements when old ones are damaged, etc. |
| | | 54 | Provision for feeding cattle. |
| | | 107 | Soil conservation. |

| B2 (3 in common) | | rpg Set 5 (paired with B2) | |
|---|---|---|---|
| 30 | Efficient and rapid distribution of seeds, fertilizer, etc., from block HQ. | 30 | Efficient and rapid distribution of seeds, fertilizer, etc., from block HQ. |
| 61 | Sufficient fluid employment for laborers temporarily (seasonally) out of work. | 61 | Sufficient fluid employment for laborers temporarily (seasonally) out of work. |
| 97 | Minimize transportation costs for bulk produce (grain, potatoes, etc.). | 97 | Minimize transportation costs for bulk produce (grain, potatoes, etc.). |
| 35 | Protection of crops from insects, weeds, disease. | 84 | Accommodation for panchayat records, meetings, etc. |
| 46 | Respect for traditional agricultural practices. | 96 | Access to railway station. |
| 47 | Need for new implements when old ones are damaged, etc. | 120 | Curative measures for disease available to villagers. |
| 98 | Daily produce requires cheap and constant (monsoon) access to market. | 121 | Facilities for birth, pre- and post-natal care, birth control. |
| | | 127 | Contact with block development officer and extension officers. |
| | | 132 | Need to develop projects which benefit from government subsidies. |
| | | 139 | Proper connection with bridges, roads, hospitals, schools, |

| B3 (3 in common) | | rpg Set 6 (paired with B3) | |
|---|---|---|---|
| 42 | Efficient plowing, weeding, harvesting, leveling. | 42 | Efficient plowing, weeding, harvesting, leveling. |
| 43 | Consolidation of land. | 43 | Consolidation of land. |
| 69 | Fullest possible irrigation benefit derived from available water. | 69 | Fullest possible irrigation benefit derived from available water. |
| 18 | Need to divide land among sons of successive generations. | 32 | Reclamation and use of uncultivated land. |
| 19 | People want to own land personally. | 39 | Best cotton and cash crop. |
| 22 | Abolition of Zamindari and uneven land distribution. | 48 | Scarcity of land. |
| 28 | Proper boundaries of ownership and maintenance responsibility. | 75 | Drainage of land to prevent waterlogging, etc. |
| 33 | Fertile land to be used to best advantage. | 104 | Plant ecology to be kept healthy. |
| 49 | Cooperative farming. | | |
| 74 | Maintenance of irrigation facilities. | | |
| 107 | Soil conservation. | | |
| 110 | Prevent land erosion. | | |

| B4 (6 in common) | | rpg Set 7 (paired with B4) | |
|---|---|---|---|
| 45 | Development of horticulture. | 45 | Development of horticulture. |
| 70 | Full collection of underground water for irrigation. | 70 | Full collection of underground water for irrigation. |
| 71 | Full collection of monsoon water for use. | 71 | Full collection of monsoon water for use. |
| 73 | Conservation of water resources for future. | 73 | Conservation of water resources for future. |
| 105 | Insufficient forest land. | 105 | Insufficient forest land. |
| 109 | Reclamation of eroded land, gullies, etc. | 109 | Reclamation of eroded land, gullies, etc. |
| 32 | Reclamation and use of uncultivated land. | 74 | Maintenance of irrigation facilities. |
| 48 | Scarcity of land. | 110 | Prevent land erosion. |
| 75 | Drainage of land to prevent waterlogging, etc. | | |
| 104 | Plant ecology to be kept healthy. | | |
| 108 | Road and dwelling erosion. | | |

## Notes on "Notes on the Synthesis of Form"

| C1 (7 in common) | | rpg Set 8 (paired with C1) | |
|---|---|---|---|
| 11 | Marriage is to person from another village. | 11 | Marriage is to person from another village. |
| 64 | Simplify the mobility of labor, to and from villages, and to and from fields and industries and houses. | 64 | Simplify the mobility of labor, to and from villages, and to and from fields and industries and houses. |
| 95 | Access to bus as near as possible. | 95 | Access to bus as near as possible. |
| 100 | Provision for bicycle age in every village by 1965. | 100 | Provision for bicycle age in every village by 1965. |
| 112 | Access to a secondary school. | 112 | Access to a secondary school. |
| 133 | Social integration with neighboring villages. | 133 | Social integration with neighboring villages. |
| 134 | Wish to keep up with achievements of neighboring villages. | 134 | Wish to keep up with achievements of neighboring villages. |
| 8 | Members of castes maintain their caste profession as far as possible. | 126 | Close contact with village-level worker. |
| 10 | Need for elaborate weddings. | | |
| 14 | Economic integration of village on payment-in-kind basis. | | |
| 15 | Modern move toward payment in cash. | | |
| 58 | Development of other animal industry. | | |
| 63 | Development of village industry. | | |
| 65 | Diversification of villages' economic base—not all occupations agricultural. | | |
| 66 | Efficient provision and use of power. | | |
| 93 | Lighting. | | |
| 96 | Access to railway station. | | |
| 99 | Industry requires strong transportation support. | | |
| 121 | Facilities for birth, pre- and post-natal care, birth control. | | |
| 130 | Need for increased incentives and aspirations. | | |
| 132 | Need to develop projects which benefit from government subsidies. | | |
| 139 | Proper connection with bridges, roads, hospitals, schools, | | |
| 141 | Prevent migration of young people and harijans to cities. | | |

| C2 (7 in common) | | rpg Set 9 (paired with C2) | |
|---|---|---|---|
| 6 | Wish for temples. | 6 | Wish for temples. |
| 21 | Eradication of untouchability. | 21 | Eradication of untouchability. |
| 111 | Provision for primary education. | 111 | Provision for primary education. |
| 115 | Opportunity for youth activities. | 115 | Opportunity for youth activities. |
| 116 | Improvement of adult literacy. | 116 | Improvement of adult literacy. |
| 135 | Spread of official information about taxes, elections, etc. | 135 | Spread of official information about taxes, elections, etc. |
| 137 | Radio communication. | 137 | Radio communication. |
| 5 | Provision for festivals and religious meetings. | 23 | Men's groups chatting, smoking, even late at night. |
| 20 | People of different factions prefer to have no contact. | 82 | Provision for children to play (under supervision). |
| 24 | Place for village events—dancing, plays, singing, etc., wrestling. | | |
| 84 | Accommodation for panchayat records, meetings, etc. | | |
| 89 | Provision of goods, for sale. | | |
| 102 | Accommodation for processions. | | |
| 117 | Spread of information about birth control, disease, etc. | | |
| 120 | Curative measures for disease available to villagers. | | |
| 129 | Factions refuse to cooperate or agree. | | |
| 140 | Develop rural community spirit: destroy selfishness, isolationism. | | |

| D1 (8 in common) | | rpg Set 10 (paired with D1) | |
|---|---|---|---|
| 26 | Sentimental system: wish not to destroy old way of life; love of present habits governing bathing, food, etc. | 26 | Sentimental system: wish not to destroy old way of life; love of present habits governing bathing, food, etc. |
| 29 | Provision for daily bath, segregated by sex, caste, and age. | 29 | Provision for daily bath, segregated by sex, caste, and age. |
| 56 | Sheltered accommodation for cattle (sleeping, milking, feeding). | 56 | Sheltered accommodation for cattle (sleeping, milking, feeding). |
| 67 | Drinking water to be good, sweet. | 67 | Drinking water to be good, sweet. |
| 92 | House has to be cleaned, washed, drained. | 92 | House has to be cleaned, washed, drained. |
| 122 | Disposal of human excreta. | 122 | Disposal of human excreta. |
| 123 | Prevent breeding germs and disease starters. | 123 | Prevent breeding germs and disease starters. |
| 124 | Prevent spread of human disease by carriers, infection, contagion. | 124 | Prevent spread of human disease by carriers, infection, contagion. |
| 76 | Flood control to protect houses, roads, etc. | 34 | Full collection of natural manure (animal and human). |
| 85 | Everyone's accommodation for sitting and sleeping should be protected from rain. | 55 | Cattle access to water. |
| 87 | Safe storage of goods. | | |
| 90 | Better provision for preparing meals. | | |

# Notes on "Notes on the Synthesis of Form"

| D2 (5 in common) | | rpg Set 11 (paired with D2) | |
|---|---|---|---|
| 1 | Harijans regarded as ritually impure, untouchable, etc. | 1 | Harijans regarded as ritually impure, untouchable, etc. |
| 9 | Members of one caste like to be together and separate from others, and will not eat or drink together. | 9 | Members of one caste like to be together and separate from others, and will not eat or drink together. |
| 13 | Family solidarity and neighborliness even after separation. | 13 | Family solidarity and neighborliness even after separation. |
| 68 | Easy access to drinking water. | 68 | Easy access to drinking water. |
| 86 | No overcrowding. | 86 | No overcrowding. |
| 12 | Extended family is in one house. | 17 | Village has fixed men's social groups. |
| 25 | Assistance for physically handicapped, aged, widows. | 49 | Cooperative farming. |
| 27 | Family is authoritarian. | 83 | In summer people sleep in open. |
| 62 | Provision of cottage industry and artisan workshops and training. | 101 | Pedestrian traffic within village. |
| 81 | Security for women and children. | 129 | Factions refuse to cooperate or agree. |
| 113 | Good attendance in school. | | |
| 114 | Development of women's independent activities. | | |

| D3 (6 in common) | | rpg Set 12 (paired with D3) | |
|---|---|---|---|
| 2 | Proper disposal of dead. | 2 | Proper disposal of dead. |
| 3 | Rules about house door not facing south. | 3 | Rules about house door not facing south. |
| 4 | Certain water and certain trees are thought of as sacred. | 4 | Certain water and certain trees are thought of as sacred. |
| 78 | Shade for sitting and walking. | 78 | Shade for sitting and walking. |
| 79 | Provision of cool breeze. | 79 | Provision of cool breeze. |
| 88 | Place to wash and dry clothes. | 88 | Place to wash and dry clothes. |
| 16 | Women gossip extensively while bathing, fetching water, on way to field latrines, etc. | 38 | Provision of threshing floor and its protection from marauders. |
| 17 | Village has fixed men's social groups. | 50 | Protected storage of fodder. |
| 23 | Men's groups chatting, smoking, even late at night. | 52 | Improve quantity of fodder available. |
| 82 | Provision for children to play (under supervision). | 77 | Village and individual houses must be protected from fire. |
| 83 | In summer people sleep in open. | | |
| 101 | Pedestrian traffic within village. | | |
| 119 | Efficient use of school; no distraction of students. | | |

| Unpaired rpg Set 13 | |
|---|---|
| 5 | Provision for festivals and religious meetings. |
| 10 | Need for elaborate weddings. |
| 20 | People of different factions prefer to have no contact. |
| 24 | Place for village events—dancing, plays, singing, etc., wrestling. |
| 37 | Provision of storage for distributing and marketing crops. |
| 102 | Accommodation for processions. |
| 128 | Price assurance for crops. |
| 131 | Panchayat must have more power and respect. |
| 140 | Develop rural community spirit: destroy selfishness, isolationism. |

| Unpaired rpg Set 14 | |
|---|---|
| 14 | Economic integration of village on payment-in-kind basis. |
| 62 | Provision of cottage industry and artisan workshops and training. |
| 63 | Development of village industry. |
| 66 | Efficient provision and use of power. |
| 89 | Provision of goods, for sale. |
| 93 | Lighting. |
| 130 | Need for increased incentives and aspirations. |
| 138 | Achieve economic independence so as not to strain national transportation and resources. |
| 141 | Prevent migration of young people and harijans to cities. |

| Unpaired rpg Set 15 | |
|---|---|
| 16 | Women gossip extensively while bathing, fetching water, on way to field latrines, etc. |
| 25 | Assistance for physically handicapped, aged, widows. |
| 27 | Family is authoritarian. |
| 81 | Security for women and children. |
| 113 | Good attendance in school. |
| 114 | Development of women's independent activities. |
| 117 | Spread of information about birth control, disease, etc. |
| 119 | Efficient use of school; no distraction of students. |

| Unpaired rpg Set 16 | |
|---|---|
| 18 | Need to divide land among sons of successive generations. |
| 22 | Abolition of Zamindari and uneven land distribution. |
| 31 | Efficient distribution of fertilizer, manure, seed, from village storage to fields. |
| 44 | Crops must be brought home from fields. |
| 98 | Daily produce requires cheap and constant (monsoon) access to market. |

## K  Cohesion and Coupling in Three Examples

In Appendix J we looked at the requirements texts associated with Alexander's decomposition of the Indian Village and with two my programs did: the first is simply his program but searching many more starting partitions, and the second uses my goodness measure, HIDECS2-rpg. In this Appendix we'll look at the partitions a different, more numeric way.

For the three decompositions I computed the strengths of cohesion and coupling. The results are in three tables: for Alexander's decomposition as reported in "Notes" see Figure 18; for the table for *Entire Village (rpg₁)* (Figure 16) see Figure 19; and for the table for *Entire Village (rpg₂)* (Figure 17) see Figure 20.

The cohesion measures are along the diagonal and coupling is off the diagonal. For cohesion I show the ratio of the number of interaction links that are entirely within a partition set to the theoretical maximum, which is $\frac{|M|\cdot(|M|-1)}{2}$, where $M$ is the partition set. Larger numbers are better, and the range is $[0, 1]$.

**Notes on "Notes on the Synthesis of Form"**

|    | A1   | A2   | A3   | B1   | B2   | B3   | B4   | C1   | C2   | D1   | D2   | D3   |
|----|------|------|------|------|------|------|------|------|------|------|------|------|
| A1 | .444 | .173 | .095 | .210 | .063 | .056 | .061 | .092 | .052 | .074 | .037 | .009 |
| A2 | .173 | .778 | .238 | .123 | .079 | .231 | .131 | .029 | .046 | .111 | .176 | .077 |
| A3 | .095 | .238 | .381 | .063 | .061 | .179 | .052 | .037 | .025 | .155 | .095 | .099 |
| B1 | .210 | .123 | .063 | .500 | .222 | .231 | .172 | .116 | .098 | .065 | .083 | .017 |
| B2 | .063 | .079 | .061 | .222 | .476 | .226 | .130 | .193 | .084 | .012 | .036 | .055 |
| B3 | .056 | .231 | .179 | .231 | .226 | .682 | .341 | .058 | .059 | .076 | .125 | .051 |
| B4 | .061 | .131 | .052 | .172 | .130 | .341 | .618 | .087 | .032 | .068 | .061 | .133 |
| C1 | .092 | .029 | .037 | .116 | .193 | .058 | .087 | .447 | .184 | .058 | .174 | .064 |
| C2 | .052 | .046 | .025 | .098 | .084 | .059 | .032 | .184 | .353 | .034 | .142 | .113 |
| D1 | .074 | .111 | .155 | .065 | .012 | .076 | .068 | .058 | .034 | .576 | .292 | .141 |
| D2 | .037 | .176 | .095 | .083 | .036 | .125 | .061 | .174 | .142 | .292 | .576 | .205 |
| D3 | .009 | .077 | .099 | .017 | .055 | .051 | .133 | .064 | .113 | .141 | .205 | .385 |

■ **Figure 18** Cohesion and Coupling Strengths for Alexander's Decomposition

|    | 1    | 2    | 3    | 4    | 5    | 6    | 7    | 8    | 9    | 10   | 11   | 12   | 13   | 14   | 15   | 16   |
|----|------|------|------|------|------|------|------|------|------|------|------|------|------|------|------|------|
| 1  | .709 | .258 | .222 | .182 | .149 | .045 | .057 | .143 | .136 | .045 | .025 | .125 | .000 | .167 | .091 | .023 |
| 2  | .258 | .400 | .037 | .212 | .060 | .083 | .167 | .143 | .042 | .028 | .065 | .021 | .056 | .028 | .000 | .042 |
| 3  | .222 | .037 | .528 | .222 | .040 | .157 | .056 | .032 | .042 | .000 | .043 | .056 | .000 | .074 | .111 | .042 |
| 4  | .182 | .212 | .222 | .564 | .104 | .144 | .125 | .104 | .159 | .061 | .146 | .068 | .030 | .106 | .152 | .125 |
| 5  | .149 | .060 | .040 | .104 | .571 | .304 | .259 | .133 | .027 | .155 | .071 | .152 | .190 | .048 | .024 | .036 |
| 6  | .045 | .083 | .157 | .144 | .304 | .682 | .188 | .119 | .042 | .167 | .120 | .083 | .056 | .014 | .000 | .052 |
| 7  | .057 | .167 | .056 | .125 | .259 | .188 | .643 | .286 | .078 | .146 | .062 | .172 | .083 | .042 | .021 | .062 |
| 8  | .143 | .143 | .032 | .104 | .133 | .119 | .286 | .667 | .071 | .000 | .032 | .089 | .000 | .071 | .024 | .036 |
| 9  | .136 | .042 | .042 | .159 | .027 | .042 | .078 | .071 | .536 | .188 | .215 | .172 | .042 | .083 | .104 | .125 |
| 10 | .045 | .028 | .000 | .061 | .155 | .167 | .146 | .000 | .188 | .467 | .194 | .125 | .000 | .028 | .028 | .083 |
| 11 | .025 | .065 | .043 | .146 | .071 | .120 | .062 | .032 | .215 | .194 | .601 | .312 | .130 | .130 | .148 | .139 |
| 12 | .125 | .021 | .056 | .068 | .152 | .083 | .172 | .089 | .172 | .125 | .312 | .571 | .125 | .167 | .104 | .156 |
| 13 | .000 | .056 | .000 | .030 | .190 | .056 | .083 | .000 | .042 | .000 | .130 | .125 | .333 | .000 | .056 | .042 |
| 14 | .167 | .028 | .074 | .106 | .048 | .014 | .042 | .071 | .083 | .028 | .130 | .167 | .000 | .333 | .167 | .021 |
| 15 | .091 | .000 | .111 | .152 | .024 | .000 | .021 | .024 | .104 | .028 | .148 | .104 | .056 | .167 | .667 | .208 |
| 16 | .023 | .042 | .042 | .125 | .036 | .052 | .062 | .036 | .125 | .083 | .139 | .156 | .042 | .021 | .208 | .536 |

■ **Figure 19** Cohesion and Coupling Strengths for *Entire Village (rpg₁)* (Figure 16)

For coupling I show the ratio of the number of links between the two partition sets to the theoretical maximum, which is $|M| \cdot |N|$. Smaller numbers are better, and the range is $[0, 1]$.

By looking at the tables we can see that the diagonals have generally larger values than the off-diagonals; this goes hand-in-hand with the goal of strong cohesion with weak coupling. Staring at numbers, though, can be decisive but mind bending. Two plots show the data in a more intuitive way. One shows cohesion and the other coupling.

Each of them takes the associated measured ratios—actual to theoretical maximum—for each trait (cohesion, coupling), sorts them from high to low, and plots those points connected by straight lines. The plot for cohesion is in Figure 21. "CA" refers to the decomposition of the Indian Village in "Notes"; "rpg1" to *Entire Village (rpg₁)* (Figure 16); and "rpg2" to *Entire Village (rpg₂)* (Figure 17). Here we can see that Alexander's decomposition fairly consistently exhibits lower cohesion than rpg1 and rpg2. This is because of two things: first, the goodness measure is looking for weak coupling, which generally goes with stronger cohesion, and second, both rpg1 and rpg2 perform more thorough searches.

|    | 1    | 2    | 3    | 4    | 5    | 6    | 7    | 8    | 9    | 10   | 11   | 12   | 13   | 14   | 15   | 16   |
|----|------|------|------|------|------|------|------|------|------|------|------|------|------|------|------|------|
| 1  | .571 | .250 | .172 | .222 | .125 | .125 | .141 | .125 | .089 | .000 | .012 | .056 | .038 | .050 | .031 | .062 |
| 2  | .250 | .533 | .225 | .189 | .089 | .167 | .088 | .090 | .114 | .040 | .030 | .033 | .100 | .280 | .112 | .062 |
| 3  | .172 | .225 | .679 | .389 | .125 | .153 | .078 | .088 | .018 | .025 | .038 | .111 | .077 | .225 | .062 | .094 |
| 4  | .222 | .189 | .389 | .722 | .148 | .210 | .111 | .200 | .127 | .033 | .044 | .025 | .111 | .089 | .083 | .125 |
| 5  | .125 | .089 | .125 | .148 | .472 | .210 | .028 | .244 | .063 | .033 | .067 | .049 | .077 | .156 | .028 | .042 |
| 6  | .125 | .167 | .153 | .210 | .210 | .556 | .181 | .111 | .016 | .033 | .067 | .037 | .026 | .022 | .056 | .028 |
| 7  | .141 | .088 | .078 | .111 | .028 | .181 | .607 | .275 | .143 | .225 | .075 | .056 | .019 | .025 | .016 | .016 |
| 8  | .125 | .090 | .088 | .200 | .244 | .111 | .275 | .622 | .157 | .180 | .130 | .144 | .138 | .080 | .112 | .138 |
| 9  | .089 | .114 | .018 | .127 | .063 | .016 | .143 | .157 | .714 | .286 | .157 | .079 | .088 | .114 | .125 | .036 |
| 10 | .000 | .040 | .025 | .033 | .033 | .033 | .225 | .180 | .286 | .667 | .120 | .144 | .146 | .060 | .050 | .062 |
| 11 | .012 | .030 | .038 | .044 | .067 | .067 | .075 | .130 | .157 | .120 | .378 | .122 | .085 | .140 | .188 | .112 |
| 12 | .056 | .033 | .111 | .025 | .049 | .037 | .056 | .144 | .079 | .144 | .122 | .500 | .145 | .222 | .056 | .042 |
| 13 | .038 | .100 | .077 | .111 | .077 | .026 | .019 | .138 | .088 | .146 | .085 | .145 | .474 | .292 | .202 | .288 |
| 14 | .050 | .280 | .225 | .089 | .156 | .022 | .025 | .080 | .114 | .060 | .140 | .222 | .292 | .900 | .275 | .275 |
| 15 | .031 | .112 | .062 | .083 | .028 | .056 | .016 | .112 | .125 | .050 | .188 | .056 | .202 | .275 | .750 | .422 |
| 16 | .062 | .062 | .094 | .125 | .042 | .028 | .016 | .138 | .036 | .062 | .112 | .042 | .288 | .275 | .422 | .750 |

■ **Figure 20** Cohesion and Coupling Strengths for *Entire Village (rpg₂)* (Figure 17)

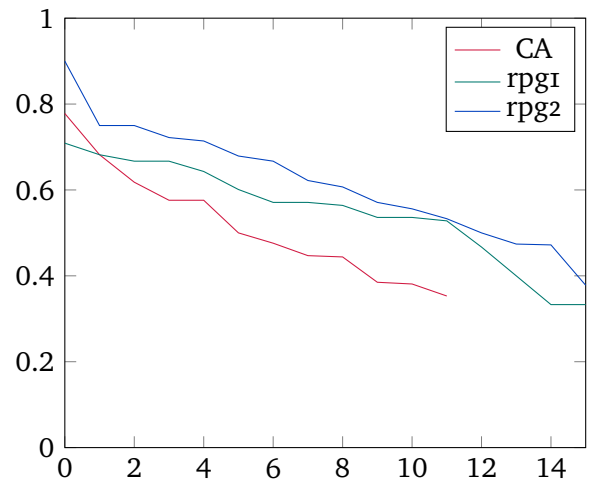

■ **Figure 21** Cohesion Graph

Now let's look at coupling: Figure 22. Here we see that Alexander's decomposition has better coupling overall. That rpg1 measures out as better than Alexander's according to `HIDECS2-Notes` seems to indicate that `HIDECS2-Notes` is not well behaved when comparing partitions of different sizes. That we cannot see Alexander's complete decomposition makes speculation difficult.

## L  A Modern Approach

I found a modern approach to the same basic problem Alexander tried to tackle in the work of Newman and Girvan in 2004 [26, 27]. They write:

> *We propose and study a set of algorithms for discovering community structure in networks—natural divisions of network nodes into densely connected subgroups.... We also propose a measure for the strength of the community structure*

**Notes on "Notes on the Synthesis of Form"**

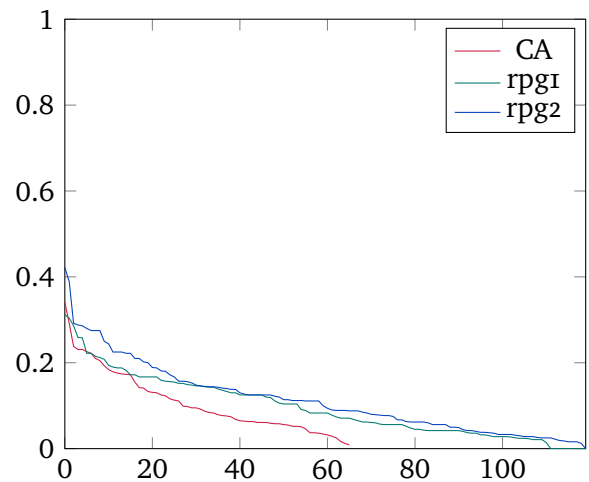

**Figure 22**  Coupling Graph

*found by our algorithms, which gives us an objective metric for choosing the number of communities into which a network should be divided.*

Their goodness measure is easy to express but a little tricky to compute. Assume that the set of nodes has been partitioned into *n* sets (or *communities*, as they call them):

$$e_{ij} = \text{the fraction of all edges linking vertices in community } i \text{ to vertices in community } j \tag{70}$$

$$a_i = \sum_j e_{ij} \tag{71}$$

$$Q = \sum_i (e_{ii} - a_i^2) \tag{72}$$

The tricky part is that each edge should contribute only to $e_{ij}$ once, either above or below the diagonal, but not both. For example, one can split the contribution of each edge half-and-half between $e_{ij}$ and $e_{ji}$, except for those edges that join a group to itself, whose contribution belongs entirely to the single diagonal element $e_{ii}$.

First let's take a look at how this measure rates the three decompositions we've been looking at in Appendices I, J, and K: the Alexander decomposition of the Indian Village problem, *Entire Village (rpg₁)* (Figure 16), and *Entire Village (rpg₂)* (Figure 17).

| Decomp | Q |
|---|---|
| CA | .208 |
| rpg1 | .193 |
| rpg2 | .176 |

Larger values are better, but Newman states that anything below .3 exhibits weak community structure. Notice that Alexander's is the best. But is it good?

Newman suggests an "agglomeration" algorithm using Q as the goodness measure; his algorithm is essentially BLDUP as suggested by Alexander in HIDECS 3. I decided to use STABL instead, which Alexander quickly moved to from BLDUP. The result is interesting. Newman mentions that Q with an agglomeration algorithm can determine the best number of communities. In this way it is similar to Karger's algorithm. Here is the result:

Set 1: 1, 2, 3, 4, 7, 9, 12, 13, 16, 17, 25, 26, 27, 29, 34, 55, 56, 67, 68, 76, 77, 78, 79, 80, 81, 83, 85, 86, 87, 88, 90, 91, 92, 94, 101, 103, 108, 113, 114, 119, 122, 123, 124, 136

Set 2: 5, 6, 8, 10, 11, 14, 15, 20, 21, 23, 24, 30, 58, 59, 61, 62, 63, 64, 65, 66, 72, 82, 84, 89, 93, 95, 96, 99, 100, 102, 111, 112, 115, 116, 117, 120, 121, 125, 127, 128, 129, 130, 131, 132, 133, 134, 135, 137, 138, 139, 140, 141

Set 3: 18, 19, 22, 28, 31, 32, 33, 35, 36, 37, 38, 39, 40, 41, 42, 43, 44, 45, 46, 47, 48, 49, 50, 51, 52, 54, 60, 69, 70, 71, 73, 74, 75, 97, 98, 104, 105, 106, 107, 109, 110, 118

Set 4: 53, 57, 126

which measures at Q=.297, which is nearly the .3 threshold Newman suggests. Note there are four sets, not 12 nor 16.

For HIDECS3-Graph, this program produces this:

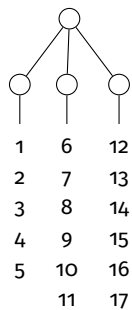

which is the same as my programs; it measures at Q=.472, which is a respectable community structure.

As with Alexander's programs, this one does better with networks that represent actual communities.

A final quote from Newman and Girvan, both physicists:

> ...*we now define a measure of the quality of a particular division of a network, which we call the* modularity. [27]

## M  Graph A Input

Errors in Alexander's data and how his programs treated typos (*punchos?*) can be better appreciated by looking at one of the examples in the HIDECS 2 report. It is **Graph A** on page D3:

**Notes on "Notes on the Synthesis of Form"**

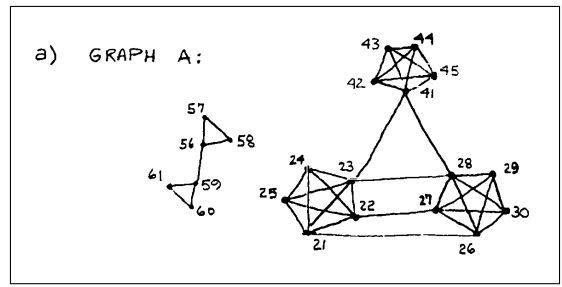

Figure 24 is a printout of the input cards for the program (on page D4 of the HIDECS 2 report). When you see it, you will think—as I did—about Alexander's handling of typos: Oh.

The HIDECS 2 program with all three of the following goodness measures partitioned **Graph A** the same:

1. HIDECS2-Actual
2. HIDECS2-Decomp
3. HIDECS2-rpg

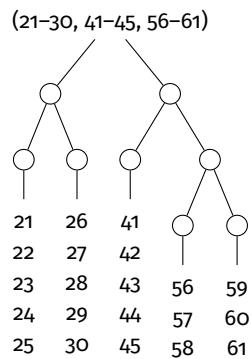

which is exactly what Alexander says his program produced (see Figure 23).

When you look at the raw input in Figure 24, notice the light pencil traces where someone has tried to decode the placement of 1s to relate them to the diagram of **Graph A**.

### N  HIDECS 2, page 25

This page (see Figure 25) is the only place where the computation of HIDECS2-Actual (see Appendix D and Figure 7) is shown—it is the only test case I could find aside from trying to replicate decompositions.

### O  Failing to Understand HIDECS2-Actual

There is at least one curiosity in the goodness measure HIDECS2-Actual (Figure 7), at least for me; perhaps it is my own mathematical unsophistication. If VSET is the set of

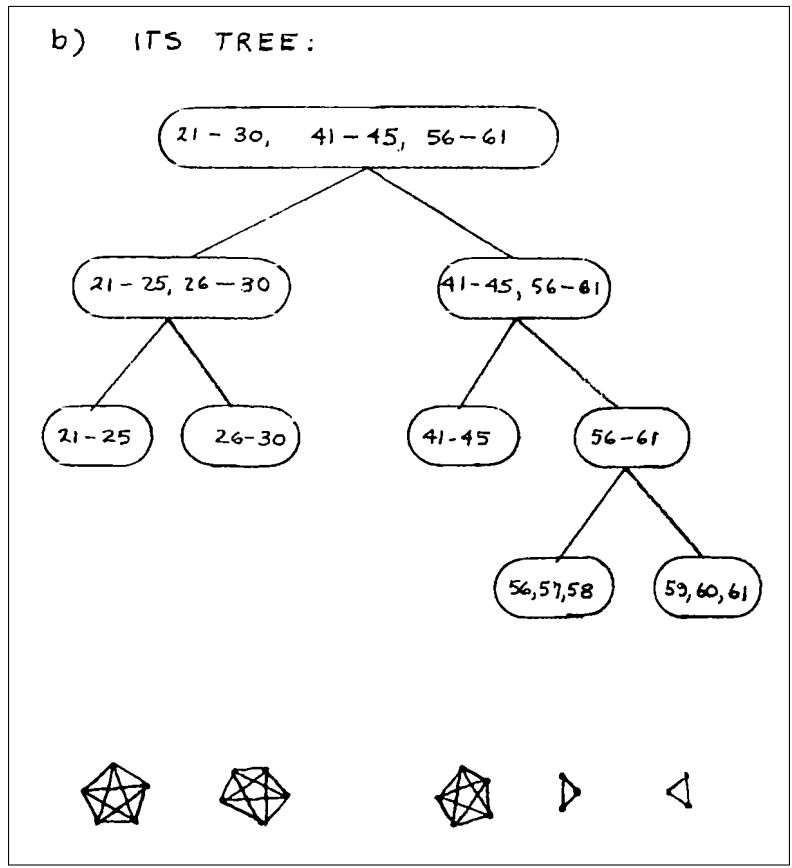

■ **Figure 23** Alexander's Tree for **Graph A**

all nodes being partitioned, and $[M, N]$ is a partition (that is, VSET $= M \cup N$), then $RR$ is the number of interaction links that cross the $M/N$ partition boundary. Alexander puts it this way in HIDECS 2 (page E2):

> *We normalize the measure by subtracting the expected value of RR and dividing by the square root of its variance. The normalized redundancy is....*

From this we can infer / guess that "the expected value of RR" is given by $\left(\frac{total}{nsq1}\right)mn$. In an example of computing INFO on page 25 of the HIDECS 2 report (page 25 of the HIDECS 2 report is reproduced in Appendix N) where

$$nbit = 9, total = 14, nsq1 = 36, |M| = 5$$

we have

$$\left(\frac{total}{nsq1}\right)mn = 7.7777777$$

Using a sampling technique I came up with, I estimated it to be 7.77771. But what does "its variance" mean? That is, where did he get that denominator? If we take it to mean "the variance of *RR*," the same sampling technique estimates it as 2.60, which seems reasonable for the variance. However, $mn(nsq1 - mn)$ puts it at 320. Using

# Notes on "Notes on the Synthesis of Form"

**Figure 24**   Input Cards for **Graph A**

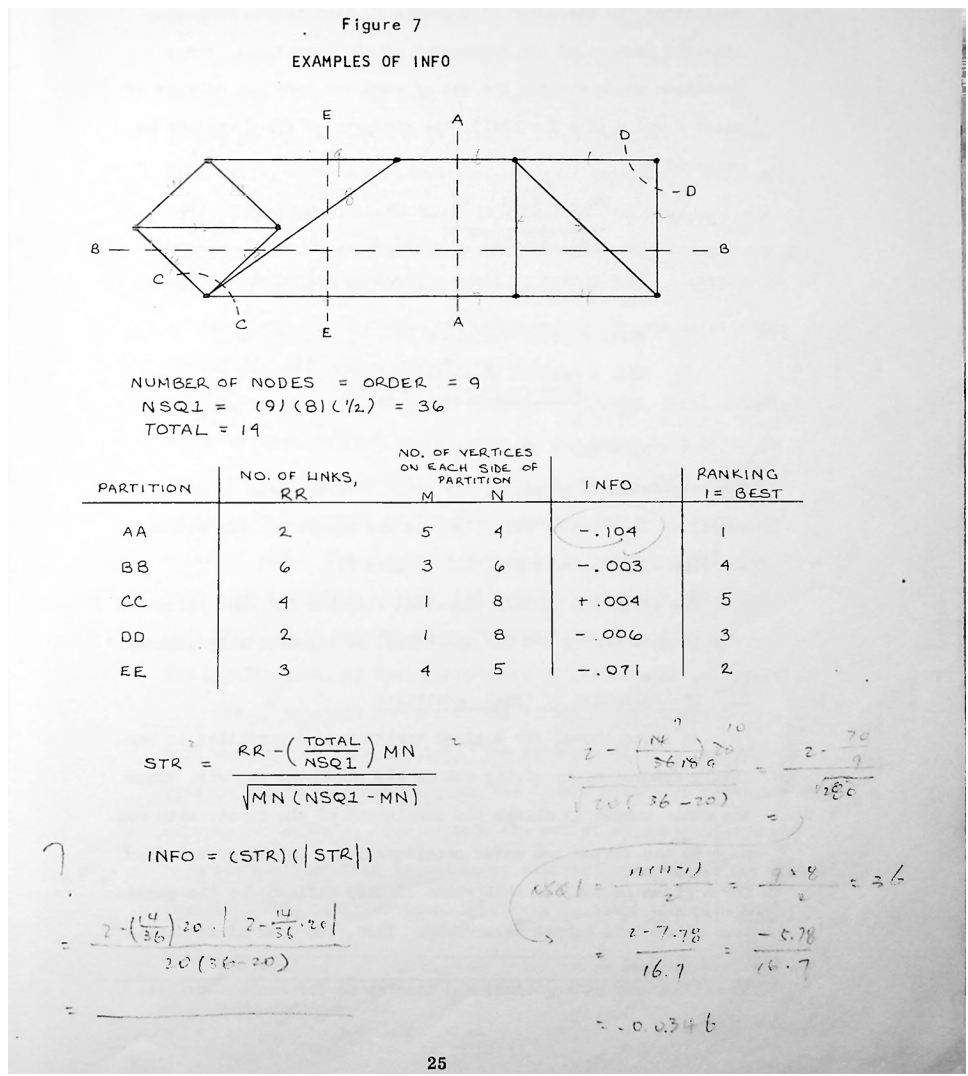

**Figure 25** HIDECS 2, page 25

the sampling technique, I also estimated the variance for the Indian Village problem: 141 nodes, 1383 total links, and a partition into a set of 75 nodes and a set of 66 nodes, $mn(nsq1-mn)$ puts it at $24,354,000$, whereas the estimate is around $293.60$.

I'm not sure $mn(nsq1-mn)$ represents the variance of anything involved in the problem, but it works well to favor balanced decompositions.

## P  Looking at Overlapping Modularity

The last example is from "Community and Privacy" [20]—written by Alexander and his collaborator at MIT, Serge Chermayeff—which contains another decomposition example; it is included in the Python bundle for HIDECS 2. The requirements and interactions are listed in Appendix Q. The Bierstone-Bernholtz recomposition for the Tomita decomposition of that set of requirements is shown in Figure 26.

**Notes on "Notes on the Synthesis of Form"**

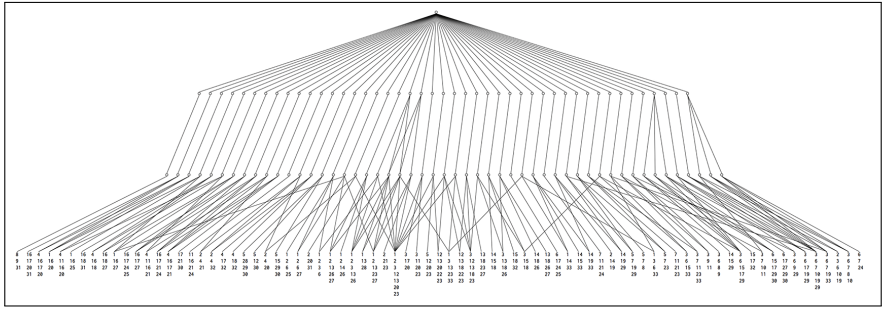

**Figure 26** Bierstone-Bernholtz Recomposition of the Tomita Decomposition for the "Community and Privacy" Problem

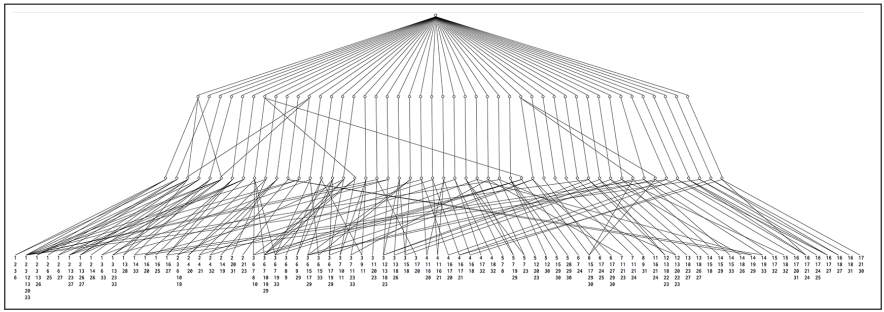

**Figure 27** Figure 26 Unoptimized

By far the most interesting part of programming Bierstone-Bernholtz recomposition was the code to minimize line crossings in the results display; the original without line-crossing optimization is shown in Figure 27.

## Q  Community and Privacy Requirements

An example that is used in the Python version of the HIDECS 2 program is the one in "Community and Privacy" [20]. I didn't do a lot with it. Here are the requirements as stated in English:

1. Efficient parking for owners and visitors; adequate maneuver space.
2. Temporary space for service and delivery vehicles.
3. Reception point to group. Sheltered delivery and waiting. Provision for information; mail, parcel, and delivery boxes; and storage of parcel carts.
4. Provision of space for maintenance and control of public utilities. Telephone, electricity, main water, sewerage, district heating, gas, air conditioning, incinerators.
5. Rest and conversation space. Children's play and supervision.
6. Private entry to dwelling, protected arrival, sheltered standing space, filter against carried dirt.
7. Congenial and ample private meeting space; washing facilities; storage for outdoor clothes and portable and wheeled objects.

8. Filters against smells, viruses, bacteria, dirt. Screens against flying insects, wind-blown dust, litter, soot, garbage.
9. Stops against crawling and climbing insects, vermin, reptiles, birds, mammals.
10. A one-way view of arriving visitors; a one-way visible access space.
11. Access points that can be securely barred.
12. Separation of children and pets from vehicles.
13. Separation of moving pedestrians from moving vehicles.
14. Protection of drivers during their transition between fast-moving traffic and the pedestrian world.
15. Arrangements to keep access clear of weather interference: overheating, wind, puddies, ice and snow.
16. Fire barriers.
17. Clear boundaries within the semi-private domain. Neighbor to neighbor; tenant to management.
18. Clear boundaries between the semi-private domain and the public domain.
19. Maintenance of adequate illumination, and absence of abrupt contrast.
20. Control at source of noises produced by servicing trucks, cars, and machinery.
21. Control at source of noises generated in the communal domain.
22. Arrangements to protect the dwelling from urban noise.
23. Arrangements to reduce urban background noise in the communal pedestrian domain.
24. Arrangements to protect the dwelling from local noise.
25. Arrangements to protect outdoor spaces from noise generated in nearby outdoor spaces.
26. Provision for unimpeded vehicular access at peak hours.
27. Provision for emergency access and escape, fire, ambulance, reconstruction, and repairs.
28. Pedestrian access from automobile to dwelling involving minimum possible distance and fatigue.
29. Pedestrian circulation without dangerous or confusing discontinuities in level or direction.
30. Safe and pleasant walking and wheeling surfaces.
31. Garbage collection point enclosed to prevent pollution of environment.
32. Efficient organisation of service intake and distribution.
33. Partial weather control between automobile and dwelling.

Here are the interactions.

1 interacts with   2, 3, 6, 12, 13, 14, 16, 20, 23, 25, 26, 27, 28, 33
2 interacts with   1, 3, 4, 6, 10, 12, 13, 14, 19, 20, 21, 23, 25, 26, 27, 31, 32
3 interacts with   1, 2, 6, 7, 8, 9, 10, 11, 12, 13, 15, 17, 18, 19, 20, 23, 26, 29, 33
4 interacts with   2, 11, 16, 17, 18, 20, 21, 32

**Notes on "Notes on the Synthesis of Form"**

5 interacts with    7, 8, 12, 15, 19, 20, 23, 25, 28, 29, 30
6 interacts with    1, 2, 3, 7, 8, 9, 10, 15, 17, 19, 24, 25, 27, 29, 30, 33
7 interacts with    3, 5, 6, 8, 10, 11, 19, 21, 23, 24, 29, 33
8 interacts with    3, 5, 6, 7, 9, 10, 31
9 interacts with    3, 6, 8, 11, 29, 31
10 interacts with    2, 3, 6, 7, 8, 11, 19, 29
11 interacts with    3, 4, 7, 9, 10, 16, 20, 21, 23, 24, 33
12 interacts with    1, 2, 3, 5, 13, 18, 20, 22, 23, 30
13 interacts with    1, 2, 3, 12, 18, 20, 22, 23, 26, 27, 28, 33
14 interacts with    1, 2, 15, 18, 19, 26, 29, 33
15 interacts with    3, 5, 6, 14, 17, 18, 29, 30, 32, 33
16 interacts with    1, 4, 11, 17, 18, 20, 21, 24, 25, 27, 31
17 interacts with    3, 4, 6, 15, 16, 19, 20, 21, 24, 25, 27, 29, 30, 31, 32
18 interacts with    3, 4, 12, 13, 14, 15, 16, 22, 23, 26, 27, 31, 32
19 interacts with    2, 3, 5, 6, 7, 10, 14, 17, 29, 33
20 interacts with    1, 2, 3, 4, 5, 11, 12, 13, 16, 17, 22, 23, 31
21 interacts with    2, 4, 7, 11, 16, 17, 23, 24, 30
22 interacts with    12, 13, 18, 20, 23
23 interacts with    1, 2, 3, 5, 7, 11, 12, 13, 18, 20, 21, 22, 27, 33
24 interacts with    6, 7, 11, 16, 17, 21, 25
25 interacts with    1, 2, 5, 6, 16, 17, 24
26 interacts with    1, 2, 3, 13, 14, 18, 27
27 interacts with    1, 2, 6, 13, 16, 17, 18, 23, 26, 29, 30
28 interacts with    1, 5, 13, 29, 30
29 interacts with    3, 5, 6, 7, 9, 10, 14, 15, 17, 19, 27, 28, 30
30 interacts with    5, 6, 12, 15, 17, 27, 28, 29
31 interacts with    2, 8, 9, 16, 17, 18, 20, 21
32 interacts with    2, 4, 15, 17, 18
33 interacts with    1, 3, 6, 7, 11, 13, 14, 15, 19, 23

As is sadly usual, these requirements have two errors, as follows:

21 interacts with:    →30, ←31

**Notes on "Notes on the Synthesis of Form"**

## About the author

**Richard P. Gabriel** is a poet, writer, and computer scientist. He lives in California. Contact him at rpg@dreamsongs.com.